\documentclass[reprint,trackchanges]{aastex631}

\usepackage{bm,amsmath}
\shorttitle{Draft}
\shortauthors{Zhang et al.}

\graphicspath{{./}{figures/}}

\begin{document}

\title{A data-driven physics-based transport model of solar energetic particles accelerated by coronal mass ejection shocks propagating through the solar coronal and heliospheric  magnetic fields}

\author[0000-0003-3529-8743]{Ming Zhang}
\affiliation{Department of Aerospace, Physics and Space Sciences, Florida Institute of Technology 
150 W. University Blvd. 
Melbourne, FL 32901, USA}

\author[0000-0002-3824-5172]{Lei Cheng}
\affiliation{Department of Aerospace, Physics and Space Sciences, Florida Institute of Technology 150 W. University Blvd. 
Melbourne, FL 32901, USA}

\author{Ju Zhang}
\altaffiliation{now at Lawrence Livermore National Laboratory, Livermore, CA 94551, USA}
\affiliation{Department of Aerospace, Physics and Space Sciences, Florida Institute of Technology 150 W. University Blvd. 
Melbourne, FL 32901, USA}

\author{Pete Riley}
\affiliation{Predictive Science Inc., 9990 Mesa Rim Rd \#170, San Diego, CA 92121, USA} 

\author{Ryun Young Kwon}
\affiliation{Korea Astronomy and Space Science Institute, Daedeokdae-ro 776, Yuseong-gu Daejeon 34055, Republic of Korea}

\author[0000-0002-3176-8704]{David Lario}
\affiliation{ NASA, Goddard Space Flight Center, Heliophysics Science Division, 8800 Greenbelt Rd. Greenbelt, MD, USA}

\author{Laura Balmaceda}
\affiliation{George Mason University, 4400 University Dr. Fairfax, Virginia 22030, USA}
\author[0000-0002-6409-2392]{Nikolai V. Pogorelov}

\affiliation{Department of Space Science and Center for Space Plasma and Aeronomic Research, University of Alabama in Huntsville, 320 Sparkman Drive, Huntsville, AL 35899, USA}

\correspondingauthor{Ming Zhang}
\email{mzhang@fit.edu}

\begin{abstract}

In an effort to develop computational tools for predicting radiation hazards from solar energetic particles (SEPs), we have created a data-driven physics-based particle transport model to calculate the injection, acceleration and propagation of SEPs from coronal mass ejection (CME) shocks traversing through the solar corona and interplanetary magnetic fields. The model runs on an input of corona and heliospheric plasma and magnetic field configuration from an MHD model driven by solar photospheric magnetic field measurements superposed with observed CME shocks determined from coronagraph images. Using several advanced computation techniques involving stochastic simulation and integration, it rigorously solves the time-dependent 5-dimensional focus transport equation in the phase space that includes pitch-angle scattering, diffusion across magnetic field line, and particle acceleration by CME shocks. We apply the model to the 2011 November 3 CME event. The calculation results reproduce multi-spacecraft SEP observations reasonably well without normalization of particle flux. This circumsolar SEP event seen by spacecraft at Earth, \textit{STEREO-A} and \textit{STEREO-B} at widely separated longitudes can be explained by diffusive shock acceleration by a single CME shock with a moderate speed.

\end{abstract}


\section{Introduction} \label{sec:intro}

When an energetic solar eruption occurs, solar energetic particles (SEPs) consisting of high-energy electrons, protons, and heavy ions up to GeV energies may be produced and released from the sun. These particles travel through the solar corona and interplanetary medium, becoming a radiation hazard to astronauts working in space and spacecraft electronics. The subject has been studied extensively for decades since energetic particle detectors can easily measure them, and analysis and modeling efforts have been made to understand them. Despite our overall understanding of their production and transport mechanisms, we still cannot reliably predict SEP radiation hazards ahead of their arrival to protect astronauts and spacecraft. The major difficulty comes from few precursory observational data and reliable models we can use to determine SEP emission and transport.  

SEPs are believed to be produced either in solar flares or at shocks driven by coronal mass ejections (CMEs). SEP events are typically classified into two classes. Impulsive events tend to be low intense events with enhanced abundances of $^{3}$He, electrons and heavy ions and are thought to be produced during magnetic reconnection processes in solar flares. These events are not a major concern of space weather because of their low radiation levels. CMEs, particularly fast ones, can drive shock waves propagating through the corona, and they sometimes can survive to a large radial distance up to many AU from the sun. CME shocks are efficient particle accelerators, picking up thermal solar wind ions or suprathermal particles from corona and interplanetary plasma material and energizing them up to several GeV. SEP events caused by CMEs are typically called gradual events, in which high levels of particle intensities, primarily protons, can last up to several days. High doses of high-energy protons are particularly dangerous to humans in space as they can penetrate deep to reach internal organs and deposit most of their energies there. Therefore, the capability to predict SEP radiation intensity and dose from CMEs will be valuable to the human endeavor of space exploration. 

CMEs are a precursor preceding the arrival of SEPs at Earth by tens of minutes to a few days. Their initiation and propagation through the corona can be remotely monitored by coronagraph instruments on spacecraft or the ground. CMEs propagate through the corona, driving shock waves visible in coronagraph images \citep[e.g.,][]{ontiveros2009}. The location, size, and speed of CME-driven shocks can be determined as early as a few tens of minutes after CME initiations \citep[e.g.,][]{kwon2014new}. The information could be used to predict SEP radiation hazards with proper models. 

Many simulation models have been published to study the acceleration and propagation of SEPs \citep{heras1992, heras1995, Kallenrode1993, Bieberetal1994, Droge1994, Ruffolo1995, KallenrodeWibberenz1997, NgReames1994, zank2000particle, giacalone2000small, Ngetal2003, Riceetal2003, Lietal2003, Lee2005, qin2006effect, zhang2009propagation, Drogeetal2010, Luhmannetal2010, Kozarevetal2013, Marshetal2015, Huetal2017, ZhangZhao2017, Li2021}. Most of these models are used to analyze SEP events and interpret underlying physics. For example, \citet{zhang2009propagation, Drogeetal2010} modeled SEP propagation by solving the Fokker-Planck transport equation with stochastic processes in a three-dimensional (3D) interplanetary magnetic field, where the idealized Parker model of the interplanetary medium is used to model the propagation of SEPs in heliospheric magnetic fields. These two models do not include the solar corona, which is thought to be the place where most of the high-energy SEPs are produced, so the production of SEPs cannot be calculated, and the effects of coronal magnetic field structures on SEP propagation cannot be quantified. Since SEP emission from the sun crucially depends on the coronal magnetic field and CME properties, both of which can vary dramatically from one solar eruption to another, prediction models of SEP radiation hazards must be based on data-driven CME propagation models under realistic coronal and heliospheric plasma and magnetic field structures. 

In this paper, we present an effort to develop a data-driven SEP model for the prediction of radiation hazards at any location in the solar system. Recently, this task has become possible thanks to the availability of sophisticated coronal and heliospheric magnetic field models based on photospheric magnetic field measurements by several helioseismic and magnetic imagers on the ground and in space (e.g., Global Oscillation Network Group operated by National Solar Observatory, and Solar Dynamics Observatory). Common methods to construct a coronal magnetic field from photospheric magnetic field measurements involve potential-field source surface, non-linear force-free field, and magnetohydrodynamics. The former two methods only consider the magnetic field, and the last method treats the plasma and magnetic fields simultaneously. The calculation of SEP acceleration and propagation requires an input of plasma and magnetic field distribution throughout its entire computation domain. Naturally, the MHD models are the best choice. Here we demonstrate how MHD models of the solar corona and heliosphere can be used to calculate SEP acceleration and propagation by a moving CME shock reconstructed from coronagraph observations. 

The data-driven model is applied to a SEP event on 2011 Nov 3 (DOY 307) observed by \textit{STEREO-A}, \textit{STEREO-B}, and \textit{SOHO} at Earth L1 point, when the three spacecraft were separated almost by 120$^\circ$ in heliocentric longitude one from another. We compute the time profiles of particle flux at various longitudes and energies. The general behaviors of SEP intensity will be investigated.

The paper is organized as follows. Section \ref{sec:model} describes the simulation model. The model contains various advanced numerical computation schemes that use stochastic integration methods to solve SEP transport equations in the phase space. Since it is the first time such methods are presented in solving high-dimensional particle transport and acceleration on vastly different scales and energies, we offer some details about the model calculation in Section \ref{sec:model}. Then, the simulation results and comparison to observations are given in Section \ref{sec:results}. Finally, Section \ref{sec:disc} presents a summary and discussion. Whereas the calculation results are not meant to provide the best fit to observations, the present study exemplifies how the model parameters determine our simulations results.

\section{Model Description} \label{sec:model}

\subsection{Governing particle transport equation}

Radiation exposure rate is measured by differential flux integrated over all energies above a given threshold. The differential flux in terms of particles per unit time, area, steradian, and energy interval is proportional to $p^2 f$, where $p$ is particle rigidity (proportional to momentum for a given charged particle species), and $f$ is the particle distribution function or phase-space density. The particle transport equation governing the evolution of energetic-particle distribution function $f(t,\mathbf{x},p,\mu)$ as a function of time $t$, position $\mathbf{x}$, $p$, and cosine of pitch angle to the outward magnetic field line $\mu$  \footnote{We choose to use the pitch angle to the outward magnetic field line as a variable instead of the regular pitch angle to the magnetic field vector because the particle distribution function and other pitch-angle-dependent quantities are expected to be more continuous upon magnetic polarity reversal at the current sheet. This is because the pitch-angle variations are mostly caused by a driver near the sun. For example, particle flux anisotropy is mainly driven by where the SEP source is, and it tends to point antisunward independent of magnetic polarity. Particle pitch angle diffusion is driven by magnetic turbulence, which is typically outward-inward anisotropic independent of magnetic polarity.} can be written as \citep[e.g.,][]{zhang2009propagation}:
\begin{eqnarray}\label{eqn:trans}
\frac{\partial f}{\partial t} {} - \nabla \cdot \bm{\kappa}_{\perp} \cdot \nabla f + \left(v \mu \hat{\mathbf{b}}_o+\mathbf{V}+\mathbf{V}_{d}\right) \cdot \nabla f - \frac{\partial}{\partial \mu} D_{\mu \mu} \frac{\partial f}{\partial \mu}+\frac{d \mu}{d t} \frac{\partial f}{\partial \mu} + \frac{d p}{d t} \frac{\partial f}{\partial p} = Q_0 (t, {\bf x}, \mu, p),
\end{eqnarray}
where the terms on the left-hand side include the following particle transport mechanisms: cross-field spatial
diffusion with a tensor $\bm{\kappa}_{\perp}$, streaming along the outward ambient magnetic field line direction $\hat{\mathbf{b}}_o$ with particle speed $v$ and
pitch-angle cosine $\mu$, convection with the background plasma velocity
$\mathbf{V}$, particle gradient\slash curvature drift velocity $\mathbf{V}_d$, pitch-angle diffusion
with a coefficient $D_{\mu \mu}$, focusing with a rate equal to $\frac{d \mu}{d t}$, adiabatic cooling with a rate
$\frac{d p}{d t}$, and on the right-hand side is the seed particle source rate $Q_0$ injected at low energies.
In the adiabatic approximation for energetic particles, the drift velocity, focusing rate, and cooling rate may be calculated from the ambient magnetic field $\mathbf{B}= B \hat{\mathbf{b}} = \pm B \hat{\mathbf{b}}_o$ ($+$ in the region of outward magnetic field polarity and $-$ in the inward magnetic field polarity) and plasma velocity $\mathbf{V}$ through
\begin{eqnarray}\label{eqn:vd}
\mathbf{V}_{d} = \frac{c p v}{q B}\left\{\frac{1-\mu^{2}}{2} \frac{\mathbf{B} \times \nabla {B}}{B^{2}}+\mu^{2} \frac{\mathbf{B} \times[(\mathbf{B} \cdot \nabla) \mathbf{B}]}{B^{3}} +\frac{1-\mu^{2}}{2} \frac{\mathbf{B}(\mathbf{B} \cdot \nabla \times \mathbf{B})}{B^{3}}\right\},
\end{eqnarray}
\begin{eqnarray}\label{eqn:focus}
\frac{d \mu}{d t} = -\frac{\left(1-\mu^{2}\right) v}{2} \hat{\mathbf{b}}_o \cdot \nabla \ln B +\frac{\mu\left(1-\mu^{2}\right)}{2} \times(\nabla \cdot \mathbf{V}-3 \hat{\mathbf{b}} \hat{\mathbf{b}}: \nabla \mathbf{V}),
\end{eqnarray}
\begin{eqnarray}\label{eqn:cooling}
\frac{d p}{d t} = -\left[\frac{1-\mu^{2}}{2}(\nabla \cdot \mathbf{V}-\hat{\mathbf{b}} \hat{\mathbf{b}}: \nabla \mathbf{V})+\mu^{2} \hat{\mathbf{b}} \hat{\mathbf{b}}: \nabla \mathbf{V}\right] p,
\end{eqnarray}
where $q$ is the particle charge. The transport equation may be accompanied by a boundary or initial condition in certain applications. We usually call Equation (\ref{eqn:trans}) the focused transport equation after \citet{Roelof1969}, although his original focused transport equation only contains particle streaming and adiabatic pitch-angle focusing along magnetic field lines. The terms in the first-order partial derivatives come from adiabatic motion of charged particles in electric and magnetic fields under the assumption of gyrotropic symmetry \citep{northrop1963adiabatic, skilling1971, isenberg1997hemispherical, qin2004interplanetary,qin2006effect, zhang2006}. 
The second-order partial derivative terms represent the effects of magnetic field
turbulence. The equation is truncated up to the diffusion term as approximated in the standard quasilinear theory \citep{zhang2006}. From the quasilinear linear theory, the diffusion tensor in the phase space should contain much more matrix elements. However, the first-order terms related to particle streaming along the magnetic field, gyration about the magnetic field, and adiabatic cooling of particle rigidity or energy are much faster than the diffusion in these variables, so the second-order derivatives of these variables can be dropped out. In particular, all the diffusion terms related to $p$ are neglected after considering that the propagation speed of magnetic field turbulence, typically the Alfv\'en speed or fast-mode MHD wave speed, is much less than the speed of particles, and stochastic particle rigidity change by electric field fluctuations in the turbulence is much slower than the adiabatic cooling during the expansion with the solar wind plasma. Furthermore, if we assume that the phase angles of magnetic field turbulence at different wavelengths are completely random, pitch-angle scattering (mainly driven by cyclotron resonance) and cross-field spatial diffusion (mainly driven by field line random walk at long wavelengths) become uncorrelated, yielding zero off-diagonal diffusion elements in the diffusion tensor \citep{jokipii1966cosmic}.

If the particle distribution is nearly isotropic, averaging the focused transport Equation (\ref{eqn:trans}) over all the pitch angles yields the Parker transport equation \citep{parker1965}. Like the Parker transport equation, the focused transport Equation (\ref{eqn:trans}) can be used to describe shock acceleration of energetic particles \citep{leRouxWebb2012, zuoetal2013a, zuoetal2013b}. Although the particle motion near the shock  is no longer adiabatic, it is a good approximation when the particle velocities are much greater than the shock speed \citep{zhang2009propagation} due to the small effect on the particles at each shock crossing. The focused transport equation allows particle distribution function to be anisotropic in pitch angle, making it more applicable than the Parker equation for many physical conditions in space including injection of seed particles at shock waves. The distribution function of SEPs near the sun or in the early phase of a SEP event could be very anisotropic, making it necessary to use the focused transport equation in modeling SEPs.

\subsection{Stochastic integration solution}

The focused transport equation is a time-dependent 5-dimensional Fokker-Planck type equation in the phase space where the gyrophase dependence is assumed to be uniform. Typical finite difference or finite element method to solve the second-order partial differential equation of this high dimension is not possible even with the fastest or largest supercomputer in the world. We use time-backward stochastic differential equations to solve it. The procedure is the following. The left-hand side of the focused transport Equation (\ref{eqn:trans}) containing all the effects of particle transport mechanisms can be rewritten with the following corresponding stochastic differential equations to describe the microphysics of particle guiding center motion and particle rigidity \citep{Gardiner1983, zhang2009propagation}:
\begin{eqnarray}\label{eqn:bk1}
d \mathbf{x}(s)&=&\sqrt{2 \kappa_{\perp}} \cdot d \mathbf{w}(s)+\left(\nabla \cdot \boldsymbol{\kappa}_{\perp}-v \mu \hat{\mathbf{b}}_o-\mathbf{V}-\mathbf{V}_d\right) d s, \\
d \mu(s)&=&\left[-\frac{d \mu}{d t}+\frac{\partial D_{\mu \mu}}{\partial \mu}\right] d s+\sqrt{2 D_{\mu \mu}} d w(s),\label{eqn:bk2} \\
d p(s)&=&-\frac{d p}{d t} d s, \label{eqn:bk3}
\end{eqnarray}
where $d w(s)$ is a Wiener process as a function of backward running time $s$. $d w(s)$ can be generated by random numbers that have a Gaussian distribution with a standard deviation of $\sqrt{ds}$.

According to \citet{Freidlin1985}, an exact solution to Equation(\ref{eqn:trans}) for any location, rigidity, pitch angle cosine and time can be written in terms of following stochastic integration and average
\begin{eqnarray}\label{solution}
f(t,{\bf x},p,\mu) = \left < \int_0^t Q_0(t-s, {\bf x}(s),p(s),\mu(s)) ds \right > + \left < f_b(t-s_e, {\bf x}_e,p_e,\mu_e) \right >
\end{eqnarray}
where $\left < \right > $ denotes the expectation value of what is inside and $f_b(t-s_e, {\bf x}_e,p_e,\mu_e)$ is the boundary or initial value of the distribution function when the backward stochastic processes described by Equations (\ref{eqn:bk1} -- \ref{eqn:bk3}) hit a boundary or the initial time for the first time (first exit point). If we choose the initial time to be before CME initiation, the inner boundary on the solar surface, and the outer boundary to be far away from the sun, $f_b$ can be set to zero. Therefore, the exact solution to the focused transport equation is just the expectation value of the integration of seed particle source rate along stochastic trajectories. We can use Monte Carlo simulations to sample important trajectories to find the integrated source. We run stochastic trajectories backward in time from the location, energy, pitch angle and time where we want to calculate the particle intensity i.e., $x(0) = x$, $\mu(0) = \mu$, and $p(0) = p$ at the initial backward time $s = 0$ at time $t$ until the CME initiation at $s=t$ or time 0. Trajectories that encounter particle sources at shock crossings will contribute to the average. Important trajectories are those that contribute significantly to the averages. Enough number of important trajectories are needed to converge the averaging calculation to a solution with a small enough statistical error bar. 

A straightforward Monte Carlo simulation with the above scheme is very inefficient. Most simulated trajectories do not encounter the shock when it is close to the sun, where the source rate is the strongest, and most particle acceleration takes place. This is because pitch-angle focusing in a radially expanding heliospheric magnetic field tends to cause the backward trajectories to go away from the sun. Furthermore, adiabatic cooling will only increase the energy in the backward simulation, thus driving the sample trajectories away from the energies of the seed particles. Very few trajectories contribute to the average with a nonzero source integration, rendering it difficult to build up the statistics needed to achieve a small enough error bar. This behavior is natural because the solution of particle distribution function far from the source is typically much lower than its value in the source region, and the probability of contributing trajectories has to be small. To overcome this problem, we have designed the following scheme to improve the sampling efficiency.

\subsection{Importance sampling with modified equation, artificial drift, killing factor and split}

Most SEPs are injected near the sun, where the seed particle density is high, the magnetic field is stronger, and the CME shock is powerful relative to those at large radial distances. The drift terms in the stochastic differential Equations (\ref{eqn:bk1}) and (\ref{eqn:bk2}) tend to drive the trajectories away from the sun, leaving most trajectories not to encounter a source at the shock and to contribute a zero value to the average. To increase the sampling efficiency of the computer simulation, we employ importance sampling, which is a Monte Carlo method for evaluating properties of a particular distribution, while only having samples generated from a different distribution than the distribution of interest \citep{kloek1978}. We modify the particle transport equation by substituting $f= (1+\mu/a) u $ with a constant tuning parameter $a>1$. The equation for $u(t, {\bf x}, \mu, p)$ becomes:
\begin{equation} \label{eqn:newtrans}
\frac{\partial u}{\partial t} - \nabla \cdot \bm{\kappa}_{\perp} \cdot \nabla u + \left(v \mu \hat{\mathbf{b}}_o+\mathbf{V}+\mathbf{V}_{d}\right) \cdot \nabla u - \frac{\partial}{\partial \mu} D_{\mu \mu} \frac{\partial u}{\partial \mu}+\frac{d \mu^\prime}{d t} \frac{\partial u}{\partial \mu} + \frac{d p}{d t} \frac{\partial u}{\partial p} = -c(t, {\bf x}, \mu, p)u + \frac{Q_0(t, {\bf x}, \mu, p)}{1 +\mu/a}.
\end{equation}
with a different rate of drift only in the pitch angle cosine:
\begin{equation} \label{eqn:newdmu}
\frac{d \mu^\prime}{dt} = \frac{d\mu}{dt} - \frac{2 D_{\mu\mu}}{a+\mu} 
\end{equation}
and a new decay or killing rate
\begin{equation} \label{eqn:killing}
c(t, {\bf x}, \mu, p) =\frac{1}{a+\mu}  \left( \frac{d\mu}{dt} - \frac{\partial D_{\mu\mu}}{\partial \mu} \right).
\end{equation}
The exact solution to the new Equation (\ref{eqn:newtrans}) can be found in \citet{Freidlin1985} or \citet{zhang2000}. The final solution to the particle distribution function can be written as:
\begin{equation} \label{eqn:newsolution}
f(t, {\bf x}, \mu, p) = (1+\mu/a) \left< \int_0^t \frac{Q_0(t-s, {\bf x}(s),p(s),\mu^\prime(s))}{1+\mu^\prime(s)/a} \exp\left( - \int_0^s c( t-s_1, {\bf x}(s_1),p(s_1),\mu^\prime(s_1)) ds_1 \right)ds \right>
\end{equation}
which is based on a different stochastic description of the pitch angle cosine
\begin{equation} \label{eqn:bk4}
d \mu^\prime(s) = \left[-\frac{d \mu^\prime}{d t}+\frac{\partial D_{\mu \mu}}{\partial \mu}\right] d s+\sqrt{2 D_{\mu \mu}} d w(s)
\end{equation}
Comparing the stochastic differential Equation (\ref{eqn:bk4}) with (\ref{eqn:bk2}), we can find that there is an additional artificial drift term of 
$\frac{2D_{\mu\mu}}{a+\mu}$ from Equation (\ref{eqn:newdmu}), which tends to drive the pitch angle cosine $\mu$ toward +1, so that the backward trajectory moves in toward the sun through Equation (\ref{eqn:bk1}). Since most seed particles of SEPs are injected near the sun, the modified stochastic trajectory tends to spend more time in the source region, raising the probability of seed source contribution. The solution in (\ref{eqn:newsolution}) contains an additional exponential factor or killing term to compensate for the increased contribution from the seed source, yielding the same answer as the solution in Equation (\ref{solution}). 

Because the new solution with the artificial drift has an increased probability of spending time in the source regions, the sampling trajectories have less chance of making zero contribution, although the exponential killing term reduces its weight to the average. In this way, the sampling trajectories make more frequent contributions with a reduced value. It makes the solution average converge more efficiently than the original method, where there are fewer regular contributions with mostly zero contributions. The speed of the artificial drift is controlled by the tuning parameter $a$. The smaller the value of $a$ is, the faster the artificial drift drives toward the sun to increase the probability of encountering the source. However, a smaller $a$ also increases the killing rate $c(t, {\bf x}, \mu, p)$. After stochastic samplings, there will be more spread in the values of the exponential killing factor, making the average more difficult to converge. A balance between these two opposing effects is needed to maximize the computer simulation efficiency in sampling important trajectories. The best choice of $a$ is difficult to quantify analytically, but numerical experimentation can help. Since the constant $a$ does not affect the result once enough statistics is achieved, we do not explicitly list the value of $a$ used in each of our calculations. 

There are other ways to increase computer simulation efficiency. For example, suppose in a simulation, we find that the stochastic trajectories have difficulty hopping across magnetic field lines to reach the seed particle source. In that case, we can design an artificial drift toward the source by properly modifying the equation. This exercise is left out of this paper.

Our code also contains a feature called trajectory split. For runs to calculate particle distribution function at a time long after the CME initiation, the important portion of sampled trajectories is near the end of backward trajectories when the CME shock is in the solar corona. The early part of the trajectory does not encounter the particle source, making this part less important. We have implemented another way of importance sampling by designing a scheme to sample more heavily towards the end. We split simulation trajectories into two after a certain amount of time. The split trajectories reduce their weight to the average by a factor of 2 for every level of splitting. The code calls the split scheme recursively until the probability runs out or the simulation end is reached. 

\subsection{Dealing with diffusive shock acceleration}

The seed particle sources in $Q_0$ have energies slightly above the solar wind energy. The energies are much lower than those of SEPs concerned as radiation hazards. For the seed particle source to contribute to the average in Equation (\ref{solution}), the simulated trajectories must go through energization in rigidity or energy via particle transport processes. Particle acceleration by the CME shock is described by the term $dp/dt$ in the transport equation (\ref{eqn:trans}) or stochastic differential Equation (\ref{eqn:bk3}). To determine how much seed particle source has been injected in the integration, we need to know the detailed processes of diffusive shock acceleration. Acceleration of low-energy particles from the seeds occurs very fast on small scales near the shock ramp. The simulation needs to trace individual shock passage back and forth. Such simulation takes a large amount of computation time, thus becoming impractical for large-scale simulation of SEP production and transport. 

We take an alternative approach to incorporate diffusive shock acceleration in this model. Particle acceleration is localized at the shock because $\frac{dp}{dt}$ is proportional to a delta function at the shock due to the spatial derivative of discontinuous plasma velocity and magnetic field. If we move the term of particle acceleration at the shock to the right-hand side of the transport equation and combine it with the seed source rate, we get a new source injection rate
\begin{eqnarray}\label{eqn:source1}
Q =Q_0 + \frac{d p}{d t}_{sh} \frac{\partial f_{sh}}{\partial p},
\end{eqnarray}
where the subscript ``sh'' denotes the quantities at any location on the shock.

Once the shock acceleration term has been moved to the source term, the gain of particle rigidity during the shock passage is no longer included in the stochastic differential equation according to the correspondence between the Fokker-Planck equation and stochastic differential equation. Note that the acceleration or cooling term away from the shock  location is still left on the left-hand side of the focused transport Equation (\ref{eqn:trans}) or in the stochastic differential Equations (\ref{eqn:bk1} - \ref{eqn:bk3}). 

Because the plasma and magnetic field properties are discontinuous at the shock, the rate of rigidity changes $\frac{dp}{dt}_{sh}$ is ambiguous primarily due to the discontinuity in the magnetic field direction relative to the shock normal. We average the shock SEP injection over all the particle pitch angles to avoid such ambiguity, assuming that the particle distribution at the shock  is isotropic. Comparison with a calculation using an anisotropic acceleration term found that the difference is minimal, probably because the particles do cross the shock in pitch angles very close to an isotropic distribution. The isotropic assumption is also expected because of the enhanced particle scattering by strong turbulence in the vicinity of a shock. So the accelerated SEP source rate can be written as follows:
\begin{eqnarray}\label{eqn:source2} 
Q = Q_0+ \frac{1}{3} \left(V_{n2}-V_{n1}\right) \delta\left(\mathbf{x}-\mathbf{x}_{s h}\right) p \frac{\partial f_{s h}(p)}{\partial p}.
\end{eqnarray}
where $V_{n1}$ and $V_{n2}$ are the upstream and downstream normal velocity component of plasma relative to the shock, respectively.

The majority of particle acceleration takes place at the shock. Without shock acceleration along the simulated backward trajectories, the particles starting at SEP high energies above 1 MeV will never reduce their energies low enough to have a significant direct contribution from the seed particle population of typically a few keV. Essentially, $Q_0$ can be considered zero, but the seed particles contribute indirectly through the injection of accelerated SEPs at the shock, which is constrained by the diffusive shock acceleration theory. $Q$ then represents the injection rate of accelerated SEP particles at the last time when they are released from the shock. In this way, we can speed up the computation and incorporate the shock acceleration without simulating the entire particle acceleration process. 

The solution to the particle transport equation can be written in the same form  as Equation (\ref{solution}) or (\ref{eqn:newsolution}) except that $Q_0$ is replaced with $Q$  and without shock acceleration in the particle trajectory simulation. The integration of the $\delta$ function in the source function $Q$ over time $\delta\left(\mathbf{x}-\mathbf{x}_{s h}\right) ds$ is called local time. We use Tanaka's formula and Ito stochastic calculus up to the second order to calculate the differential local time through the distance to shock surface for each shock crossing using the following formula \citep[see e.g.,][]{bjork2015, zhang2000}:
\begin{equation}\label{eqn:dlt}
\delta\left(\mathbf{x}-\mathbf{x}_{sh}\right) ds = \frac{d | d_{sh} | - {\rm sign}(d_{sh}) d d_{sh}}{ \boldsymbol{\kappa}_{\perp} : \hat{\mathbf{n}}_{sh} \hat{\mathbf{n}}_{sh} + [( \nabla \cdot \boldsymbol{\kappa}_{\perp}  -v \mu \hat{\mathbf{b}}_o-\mathbf{V}-\mathbf{V}_d ) \cdot \hat{\mathbf{n}}_{sh}]^2 ds /2}
\end{equation}
where $d_{sh} = \left(\mathbf{x}-\mathbf{x}_{sh}\right) \cdot \hat{\mathbf{n}}_{sh}$ is the distance to the shock surface, $\hat{\mathbf{n}}_{sh}$ is the unit vector normal to the shock, and ${\rm sign}(d_{sh})$ is the $\pm 1$ sign function of $d_{sh}$. Note that Equation(\ref{eqn:dlt}) obeys the Ito stochastic integration rule so that it is not zero only during the step when the shock is crossed. We expand stochastic calculus to the accuracy of drift speed square in case a time step is not small enough for it to be dominated by the perpendicular diffusion. In this way, the numerical integration of the $\delta$-function does not require us to approximate it with a continuous function.

The new source rate requires the particle distribution function $f_{sh}(p)$ to be known in the computer simulation. Fortunately, the particle distribution function at the shock $f_{sh}(p)$ is mostly determined by local shock conditions. It is so at least up to cut-off rigidity ($p_c$), and beyond that, the distribution function drops precipitously with the increase of $p$. The solution of particles distribution function at the shock is given by a power law with a slope $\gamma_s = 3R/(R-1)$, which is only determined by the shock compression ratio $R$ up to a cut-off moment ($p_c$) independent of the particle diffusion coefficient \citep[e.g.,][]{Drury1983} and the large-scale shock geometry. It is unlikely that SEP transport on the large-scale heliospheric magnetic field will affect the local shock acceleration of particles below the cut-off rigidity. Therefore, the particle distribution function at the shock is known as long as we know how many total seed particles have been injected at the shock. 

Time-dependent solution to diffusive shock acceleration of energetic particles can be found in \citet{Drury1983}. The solution is not in a closed analytical form for arbitrary particle diffusion coefficients, so we have adopted an approximate solution proposed by \citet{forman1983} in the following form
\begin{equation}\label{eqn:forman1983}
f_{sh}(p) = \frac{3 N}{V_{n1}-V_{n2}} \left(\frac{p}{p_{inj}} \right)^{-\gamma_s} \frac{1}{2} \left [ \exp \left( \frac{\bar{t}^2}{\delta t^2} \right) {\rm erfc} \left( \sqrt{\frac{\bar{t}^3}{2t\delta t^2}}+\sqrt{\frac{\bar{t} t}{2\delta t^2}}\right) + {\rm erfc} \left( \sqrt{\frac{\bar{t}^3}{2t\delta t^2}}-\sqrt{\frac{\bar{t} t}{2\delta t^2}}\right) \right]
\end{equation}
where $N$ is the rate of particle distribution function injected at the shock with a characteristic rigidity $p_{inj}$, $\bar{t}$ is the average acceleration time, and $\delta t^2$ is the standard deviation of particle acceleration time, which can be expressed as
\begin{equation}\label{tbar}
\bar{t}=\int_{p_{inj}}^{p} \frac{3}{V_{n1}-V_{n2}} \left [ \frac{\kappa_1(p^\prime)}{V_{n1}}+\frac{\kappa_2(p^\prime)}{V_{n2}} \right] \frac{dp^\prime}{p^\prime}
\end{equation}
\begin{equation}\label{dt2}
\delta t^2=\int_{p_{inj}}^{p} \frac{6}{V_{n1}-V_{n2}} \left [ \frac{\kappa_1^2(p^\prime)}{V_{n1}^3}+\frac{\kappa_2^2(p^\prime)}{V_{n2}^3} \right] \frac{dp^\prime}{p^\prime}
\end{equation}
with $\kappa_1$ and $\kappa_2$ being the particle diffusion coefficients upstream and downstream of the shock, respectively. The distribution is a power law as a function of $p$ up to $p_c$ determined by
\begin{equation}\label{pc}
t=\bar{t}(p_c) = \int_{p_{inj}}^{p_c} \frac{3}{V_{n1}-V_{n2}} \left [ \frac{\kappa_1(p^\prime)}{V_{n1}}+\frac{\kappa_2(p^\prime)}{V_{n2}} \right] \frac{dp^\prime}{p^\prime}
\end{equation}
where $t$ is the age of the shock since its initiation. In the presence of particle adiabatic cooling in the background solar wind, $t = {\rm min} (t, t_{cool})$, where the cooling time is $t_{cool} = 3 (\nabla \cdot {\bf V})^{-1}$. 
Because typically $\kappa_2 \ll \kappa_1$, the upstream condition essentially determines the acceleration time. We choose the Bohm limit for it or $\kappa_1 = v p /(3 q B_1)$, where $v$ is the particle speed, $q$ particle charge and $B_1$ is upstream magnetic field strength. Because of the increasing diffusion with rigidity, the acceleration time is mostly spent in the high-rigidity end.

The applicability of diffusive shock acceleration requires that the shock ramp is a sharp discontinuity. To most energetic ions above several keV, the gyroradii of these particles are much larger than the shock ramp. We expect the approach here is applicable to SEP ions. However, the shock may not behave as a discontinuity to energetic electrons up to several MeV. The acceleration of electrons should be treated as stochastic shock drift acceleration instead of diffusive shock acceleration \citep[e.g.,][]{katou2019}. The above formalism does not apply to SEP electrons.

\subsection{Coronal and heliospheric plasma and magnetic fields}

A background ambient solar wind and magnetic field throughout the entire computation domain is needed to calculate particle transport effects on the sampling trajectories. We take from the calculation result of MAS corona and CORHEL heliosphere MHD model developed by Predictive Science Inc (https://www.predsci.com/portal/home.php). The MAS/CORHEL code solves the set of resistive MHD equations in spherical coordinates on a non-uniform mesh. The details of the model have been described elsewhere \citep[e.g.,][]{mikic1994, lionello2001, riley2001, riley2011, downs2016, caplan2017}. The model is driven by the observed photospheric magnetic field. HMI magnetograph measurements \citep{scherrer2012} on the \textit{SDO} spacecraft \citep{pesnell2012} are used to construct a boundary condition for the radial magnetic field at 1 R$_\odot$ as a function of latitude and longitude. In this study, we built up a map based on observations during Carrington rotation 2116, covering the period when the SEP event on 2011 November 3 occurred. The use of a magnetic map built over a Carrington rotation implies that some of the photospheric magnetic field data may be out of date by up to 2 two weeks. The data have also been corrected for the projection effects using a pole-fitting procedure to reconstruct the magnetic field in poorly observed regions. The MAS/CORHEL model is run in two stages: first, the corona region from 1 to 30 R$_\odot$ is modeled, followed by the region from 30 R$_\odot$ to 3 AU, driven directly by the results of the coronal calculation. This approach is much more efficient computationally, and, by overlapping the region between the simulations, we verified that the transition is seamless \citep{lionello2013}. This version of the model implements a Wave-Turbulence-Driven (WTD) approach for self-consistently heating the corona and invokes the WKB approximation for wave pressures, providing the necessary acceleration of the solar wind \citep{downs2016}. It includes the physical mechanism of the solar wind heating involving the interaction of outward and reflecting Alfv\'{e}n waves and their dissipation \citep[e.g.,][]{zank1996, verdini2007}. 

We assume that the magnetic field and plasma configuration through the computational domain is stationary in a reference frame corotating with the sun for the duration of a SEP event, which could last up to a few days. A CME can dramatically disrupt the field and plasma configuration. This mainly occurs downstream of the CME shock. Below in Subsection \ref{subsecshock}, we discuss how this might affect the calculation of SEP acceleration and propagation. We compute particle trajectories in the coordinates corotating with the sun, where a tangential convection velocity component due to sun's rotation is added to the output of the MHD model run. 

Because high-energy SEPs are quite mobile in the heliospheric magnetic field, we set the outer boundary for the SEP simulation at 20 AU, i.e., a large enough radial distance where we can assume an absorptive boundary condition without affecting our calculation result. The MHD solar wind plasma and magnetic field model covers up to 3 AU. Between 3 and 20 AU, we use the line of characteristics method to extrapolate the boundary condition of plasma and magnetic field at 3 AU to larger radial distances. Our simulations found that the time-backward trajectories rarely go beyond 3 AU because the artificial drift tends to pull important trajectories toward the sun if the particle mean free path is less than 1 AU. In very rare cases, stochastic trajectories can go beyond 3 AU, but these are typically not important trajectories. As a rule of thumb, the boundary condition at locations a few times the particle mean free paths downstream will not affect what is seen by an upstream observer. Our results indicate that a 3 AU outer boundary is far enough if we want to calculate SEP intensity at 1 AU from the sun. 

\subsection{CME shock, propagation and seed source particles} \label{subsecshock}
The source of accelerated SEPs, expressed by the injection rate $Q$ in Equation (15) comoves with the CME shock. We consider the location, shape, and time propagation of the CME shock from an ellipsoid model developed by \citet{kwon2014new}. The CME shock surface is reconstructed from EUV and white-light coronagraph images taken by instruments on spacecraft such as \textit{SOHO}, \textit{SDO} and \textit{STEREO}. Many CME shocks can only be observed as a faint edge of diffuse light emission when they are in the corona up to tens of $\rm R_S$. A CME event typically contains several frames of images in which a shock can be identified so that its time evolution can be tracked. If a CME can be viewed from multiple vantage points, it is possible to unambiguously identify the CME shock 3-d geometry and its propagation. \citet{kwon2014new} modeled the CME shock as an ellipsoid one at a time from each image. Fits to the edge of diffuse light emission can yield parameters describing the ellipsoid's size, location, and orientation for the shock surface in 3D. The actual shock formed around the CME does not necessarily cover the entire ellipsoid. An additional parameter is used to specify the size of the polar angle to which the CME shock extends from the direction of the shock leading edge (front) \citep{kwon2017}. Given the surface geometry and its time evolution, we can calculate the velocity and normal vector at any point on the shock. 

Beyond the last frame, when the CME shock extends out of the field of views of the used coronagraphs, we have to rely on a model to extrapolate its propagation further into the interplanetary medium. Many CMEs exhibit slowdown after they leave the solar corona, so we cannot extrapolate the observed shock in any simple way, such as linear or quadratic extrapolation. Otherwise, we would most likely overestimate the shock speed in interplanetary space. We have adopted into our code the analytical CME propagation model suggested by \citet{corona-romero2013}, which has been tested extensively in \citet{corona-romero2017}. In this model the propagation of the CME consists of three phases. In the beginning, during the driving stage, the CME and its shock front maintain a constant speed. The first critical time ($\tau_{c1}$) indicates the time when the force from the interaction region between the CME and the ambient solar wind becomes dominant, leading to the deceleration of the CME and subsequently an increase in the stand-off distance to its shock ahead. Then a time comes when the plasma sheath has expanded so much to a level that the shock is no longer driven. The second critical time ($\tau_{c2}$) marks the transition of the shock into a blast wave, and the shock speed decreases. The evolution of CME shock speed at the shock front can be described by \citep{corona-romero2013}:
\begin{eqnarray} \label{cmeprop}
V_{shf} (t) = \left\{ \begin{array}{l@{\quad:\quad}l} V_{cme0}, & t < \tau_{c2} \\ (V_{cme0}-V_{1AU}) \left( \displaystyle \frac{t}{\tau_{c2}} \right)^{1/3} + V_{1AU}, & t < \tau_{c2} \end{array} \right. 
\end{eqnarray}
where $V_{cme0}$ is the initial speed of the CME, and $V_{1AU}$ the solar wind speed at 1 AU. The critical times can be determined by the properties of the parent solar flare and the initial CME eruption through 
\begin{equation}
\tau_{c1} = \frac{a(1+\sqrt{c})}{a-1} \Delta t_f
\end{equation}
\begin{equation}
\tau_{c2} = \frac{d_{so}}{\sqrt{V_{A2}^2+V_{S2}^2}} + \tau_{c1}
\end{equation}
with
\begin{equation}
a = \frac{V_{cme0}}{V_{1AU}} \left( \frac{1+\sqrt{c}}{\sqrt{c}} \right) - \frac{1}{\sqrt{c}},
\end{equation} 
\begin{equation}
c = \frac{n_{cme0}}{n_{1AU}} \left( \frac{r_{cme0}}{1 AU} \right)^2
\end{equation}
where $c$ is the ratio of the initial CME plasma density relative to extrapolated the solar wind density from 1 AU, $\Delta t_f$ is the duration the solar flare rise phase, $d_{so}$ is the stand-off distance between the shock and CME, and $V_{A2}$ and $V_{S2}$ are the Alfv\'{e}n and sound speed in the sheath medium at the time when the CME begins to slow down at the first critical time $\tau_{c1}$. An empirical formula combined from \citep{farris1994, bothmer1998} is used to calculate the stand-off distance:
\begin{equation}
d_{so} = 0.264 ~AU~ \left[ \frac{(\gamma-1)M_1^2+2}{(\gamma+1)(M_1^2-1)} \right] \left( \frac{r_{cme0}+V_{cme0}\tau_{c1}}{1 AU} \right)^{0.78}
\end{equation}
where $\gamma$ is the plasma adiabatic index and $M_1$ is the fast magnetosonic Mach number of the upstream plasma flow at the shock front. The post-shock Alfv\'{e}n speed $V_{A2}$ and sound speed $V_{S2}$ can be determined using shock compression calculation (see below). Integrating the CME shock speed over time yields the radial distance of the CME shock front. Once the shock front radial distance has been extrapolated from the last frame of the observed CME shock surface, we scaled the axes and radial distance of the ellipsoid proportionally so that we can calculate the entire surface. 

In most situations, the actual shock will not form at every point of the surface described by the ellipsoid, particularly on the side opposite to the direction the CME heads. We need to limit the shock size. Such size can be determined from the edge of diffuse light emission if it can be identified in coronagraph images. Beyond the last frame of the CME image, we have to use a model to extrapolate the angular size. We assume that the angular size reaches its asymptotic value after reaching 21.5 R$_\odot$. We use the published value on the Space Weather Database of Notifications, Knowledge, and Information (DONKI) website (https://kauai.ccmc.gsfc.nasa.gov/DONKI/search/). The CME size in the DONKI catalog is obtained from coronagraph measurements using NOAA Space Weather Prediction Center CME Analysis Tool \citep{pulkkinen2010, mays2015}. We use linear interpolation to determine the angular size between the last frame of observed CME shock and the time when it reaches 21.5 R$_\odot$.

We insert the partial or full ellipsoid shock surface and its time evolution into the coronal and heliospheric magnetic field and plasma model to derive the upstream shock properties at any point on the shock surface. Relevant parameters, such as shock speed relative to the solar wind plasma, shock normal, upstream magnetic obliquity, Alfv\'en Mach number, and sonic Mach number, are fed into the MHD shock adiabatic equation for the shock compression ratio $R$ calculation \citep[e.g.,][]{Thompson1962, book1987, Kabin2001JPlPh..66..259K}:
\begin{eqnarray} \label{shockadiabatic}
(1&-&R \cos^2 \theta_{bn1} M_{A1}^{-2})^2 [ (\gamma +1 - \gamma R + R) -2R M_{S1}^{-2} ] \nonumber \\ 
&&- R \sin^2 \theta_{bn1} M_{A1}^{-2} [ \gamma +(2-\gamma ) R - (\gamma +1 -\gamma R + R) R\cos^2 \theta_{bn1} M_{A1}^{-2}] =0 
\end{eqnarray}
where $\theta_{bn1}$ is the magnetic obliquity, $M_{A1} $ the Alfv\'{e}n Mach number, $M_{S1}$ the sonic Mach number of the upstream plasma flow relative to the shock normal. The shock adiabatic equation is a cubic equation. We use Vi\`{e}te's trigonometric solution to get the roots of the equation every time the shock is crossed. Plasma density $n_{sw}$, normal velocity $V_{n}$, tangential velocity $V_{t}$ relative to the shock, thermal pressure $P$, magnetic field normal $B_{n}$ and tangential $B_{t}$ components in the downstream region (denoted by the subscript 2) can be further derived from the shock compression ratio using the following equations:
\begin{eqnarray} \label{ds}
&&\frac{n_{sw2}}{n_{sw1}} = R \\
&&\frac{B_{n2}}{B_{n1}} = 1 \\
&&\frac{B_{t2}}{B_{t1}} =R \frac{1 - \cos^2 \theta_{bn1} M_{A1}^{-2}}{1 - R \cos^2 \theta_{bn1} M_{A1}^{-2}} \\
&&\frac{V_{n2}}{V_{n1}} = \frac{1}{R} \\
&&\frac{V_{t2}}{V_{t1}} =\frac{1 - \cos^2 \theta_{bn1} M_{A1}^{-2}}{1 - R \cos^2 \theta_{bn1} M_{A1}^{-2}} \\
&&\frac{P_{2}}{P_{1}} = 1 + \frac{\gamma(R-1)}{M_{S1}^{-2}R} \left[1-\frac{RM_{A1}^{-2}(\gamma+1-2RM_{A1}^{-2}\cos^2\theta_{bn1})}{2(1 - R \cos^2 \theta_{bn1} M_{A1}^{-2})^2}\right]
\end{eqnarray}
We only take the solution for a fast-mode shock with a $B_{t2}/B_{t1} > 1$. This automatically cuts out the locations where a shock cannot form.

The shock compression ratio $R$ is used to determine the slope of the accelerated SEP power-law spectrum. To assess the level of SEP source injection rate in $Q$, we need to know how many seed particles have been injected per unit time at the shock characterized by an injection rigidity $p_{inj}$. The theory of seed particle injection for shock acceleration is quite vague and still needs fundamental understanding. In reality, seed injection depends very much on the magnetic field environment and the plasma's thermal and suprathermal particle populations in the vicinity of the shock. Information about them in the solar corona is lacking, and probably their properties could vary significantly depending on solar conditions or even on the characteristics of solar events. If so, the absolute SEP intensity level calculation could suffer quite a bit of uncertainty.

We argue that most seed particles to the CME shock in the solar corona could come from thermal solar wind ions, particularly after they have been heated by the shock passage. The sonic Mach number upstream of the CME shock is not expected to be too large in the corona, mostly around a few. This means that even the upstream plasma could still contain a substantial fraction of particles in the thermal tails that can overcome the plasma convection to encounter the shock repeatedly for diffusive shock acceleration. Once they pass through a shock, they are rapidly heated to become sub-magnetosonic. Immediately downstream of the shock, the thermal tail particles could have high enough energies to overcome convection away from the shock, becoming the seed particles that can effectively participate in diffusive shock acceleration. In this simulation, we use a characteristic particle injection speed  2.4 times the shock speed, or $v_{inj}= 2.4 V_{n1}$. Then the total amount of seed particles per unit time injected at the shock can be related to the Maxwellian velocity distribution of downstream solar wind ions so that
\begin{equation}\label{totalsource}
N = \eta (\theta_{bn}) \frac{n_{sw2} V_{n1}}{(4\pi v_{th2}^2)^{3/2}} \exp\left(-\frac{v_{inj}^2}{v_{th2}^2} \right)
\end{equation}
where $\eta (\theta_{bn})$ describes the shock obliquity dependence of seed particle injection into diffusive acceleration,  and $v_{th2}$ is the downstream solar wind thermal speed, which can be determined from the plasma thermal pressure $P_2$.  We take $\eta(\theta_{bn}) = 0.8 +0.7 ~{\rm tanh} [(\theta_{bn}-60^\circ)/10^\circ]$ from a result of particle-in-cell simulation by \citet{caprioli2014}.
Because the injection speed sits in the tail of a Maxwellian distribution, the number of total injected particles is sensitive to $v_{inj}$. We found that a $v_{inj}$ between $2.3 -2.7$ $V_{n1}$ can generally produce a good fit to observations. For the 2011 November 3 event, a $v_{inj}= 2.4 V_{n1}$ turns out to be the best. With the above assumed source rate, the absolute SEP intensity can be obtained without further normalization.

In addition, the code can handle arbitrary sources of seed particles. If a particular suprathermal population is injected, we can add the total number of injected particles to Equation (\ref{totalsource}). This kind of scenario will be explored in future studies.

Since the ellipsoid shock surface and its time evolution is inserted on a steady-state plasma and magnetic field distribution without a CME eruption, the downstream plasma and magnetic field distribution inside the shock ellipsoid is not consistent with the shock jump condition. The calculation of particle acceleration would not be correct unless we modify the downstream magnetic field and plasma. This requires an input of a time-dependent plasma and magnetic field model, which will cost some computation time. However, this problem has been mitigated in our approach because the calculation of the shock acceleration process has been replaced by the injection of accelerated SEPs consistent with the diffusive shock acceleration theory and the shock compression. So we do not have to correct for the change of plasma and magnetic field due to the CME shock propagation to correctly assess the accelerated SEP source rate. Time-dependent MHD model including the propagation of CME could be implemented in future model runs.

\subsection{Diffusion coefficients}

Our model also requires an input of particle transport coefficients, such as pitch angle diffusion coefficient $D_{\mu\mu}$ and spatial diffusion perpendicular to the magnetic field $\kappa_\bot$. The magnetic field turbulence properties determine their values. There are several theories for the particle transport coefficients \citep[see recent review][]{engelbrecht2022}, but none of them have been tested rigorously. The input for calculating particle diffusion coefficients typically involves a magnetic field turbulence spectrum covering all the spatial and wavenumber domains. In addition, analysis of SEP events showed that the diffusion coefficients could change significantly from one solar event to another \citep{droge2014}. Because of these reasons, currently, the diffusion coefficients cannot be implemented as data-driven. A common practice in modeling SEPs is to treat them as free parameters until the calculation results can reasonably fit observations.

We follow an approach that we have adopted in a previous work \citep[e.g.,][]{ZhangZhao2017}, assigning
\begin{equation} \label{dmumu}
D_{\mu\mu} = D_0({\bf x}) p^{q-2} (1-\mu^2) (|\mu|^{q-1} +h_0)
\end{equation}
where the rigidity $p$ is in the unit of GV.
The expression is based on the results of the standard quasilinear theory \citep[e.g.,][]{jokipii1966, schlickeiser2002} of particle scattering by the magnetic field turbulence with a power-law spectrum of slope $-q$. We choose a Kolmogorov spectrum slope $q=5/3$ in the inertial range of wavelength. The term containing $|\mu|^{q-1}$ comes from the quasilinear resonant scattering by magnetic field fluctuations
\begin{equation} \label{dmuql}
D_{\mu\mu}^{QL} = \frac{\pi^2\Omega^2 (1-\mu^2)}{B^2v|\mu|} W_\bot(k_{res}) ~~{\rm with} ~~ k_{res}=\frac{\Omega}{v |\mu |}
\end{equation}
where $\Omega$ is the particle angular gyrofrequency and $W_\bot$ is the spectral power density of the transverse magnetic field fluctuations as a function of wavenumber $\alpha_\bot$.  We assume that the forward and backward propagating fluctuations have equal power density. The resonance condition also yields the rigidity dependence in Equation (\ref{dmumu}). The additional parameter $h_0$ is added to phenomenologically describe the enhancement of scattering through $\mu=0$ through either non-resonant scattering or non-linear effects. We set $h_0=0.2$. The result of our calculation is not very sensitive to $h_0$ unless $h_0 \ll 0.05$. 

Pitch-angle diffusion leads to a spatial diffusion of particles along magnetic field lines. The parallel mean free path can expressed as (Hasselmann and Wibberenz, 1970)
\begin{equation} \label{mfp}
\lambda_{||} = \frac{3 v} {8} \int\limits^1_{-1} d\mu \frac{(1-\mu^2)^2}{D_{\mu\mu}}  = \lambda_{||0}({\bf x}) p^{2-q} 
\end{equation}
where the rigidity $p$ is in the unit of GV and $\lambda_{||0}({\bf x})$ is the parallel mean free path at 1 GV. We generally use the value of particle mean free path to specify the intensity of particle pitch-angle scattering. The magnitude of the parallel mean free path is mainly determined by the value of $\lambda_{||0}({\bf x})$, which could be a function of location ${\bf x}$. We follow Bieber (1994) to set the radial mean free path $\lambda_r$ to be constant. Then $D_0({\bf x})$ can be determined using $\lambda_r = \lambda_{||} \cos^2\psi$, where $\psi$ is the spiral angle of the Parker magnetic field to the radial direction. 

The spatial diffusion perpendicular to the ambient magnetic field $\kappa_\bot$ could be due to the motion of particles following meandering or random-walking magnetic field lines or due to particle hopping across the ambient magnetic field by the mechanisms of turbulent drift or scattering (Jokipii, 1966). In the model runs contained in this paper, we assume $\kappa_\bot$ is driven by field line random walk started at the bottom of the solar corona. We follow the formula in \citet{ZhangZhao2017} 
\begin{equation}
\kappa_\bot = \frac{v}{2V} \alpha_\bot \kappa_{gd0} \frac{B_0}{B} 
\end{equation}
where $\kappa_{gd0} = 3.4 \times 10^{13}$ cm$^2$s$^{-1}$ is the diffusion coefficient in the photosphere estimated from a typical speed of supergranular motion, $v/V$ is the ratio of particle to solar wind plasma speed, and $B/B_0$ is the ratio of the magnetic field relative to its value on the solar surface $B_0$ on the same field line. A factor $\alpha_\bot$ is inserted to tune down the transmission of field line diffusion from the photosphere to the corona. We typically set a $\alpha_\bot$ value less than 1.

When a CME shock produces a high-enough number of energetic particles, anisotropic beam of particles propagating upstream of the shock may amplify waves through their effects on plasma instabilities. These upstream waves can act back and prevent particles from escaping the shock vicinity through pitch angle scattering. A complete SEP model should include this effect. We have only partially implemented this feature by applying the Bohm diffusion limit for calculating shock acceleration of particle sources. Still, we have not included the effect of upstream waves in the large-scale SEP propagation calculation. We assume that the upstream region affected by shock-generated plasma waves is relatively thin compared to the mean free path of the particles we simulate this paper. When a CME is not very powerful to generate a high density of SEPs near the shock, the effect of upstream waves will not be severe. The background solar wind turbulence may still be the primary driver of particle scattering. We assume that this condition could be applicable to the CME on 2011 November 3, which is not a very fast.

\section{Results} \label{sec:results}

We now apply our model calculation to the 2011 November 3 halo CME event, which caused enhancements of SEPs seen by \textit{SOHO} and \textit{ACE} at the Earth-Sun Lagrangian point L1, \textit{STEREO-A} (\textit{STA}), and \textit{STEREO-B} (\textit{STB}). It is called a circumsolar SEP event, because three spacecraft at widely separated heliographic longitudes saw SEPs shortly after the CME initiation \citep{gomez2015}. 

\subsection{Observations}

Figure 1 shows an equatorial view (top) and a projection on the solar surface (bottom) of the spacecraft's position and their connected magnetic field lines based on the MAS/CORHEL MHD model. The CME propagation direction is indicated by the purple arrows. The yellow hatched regions indicate the CME coverage of solar longitude and latitude at two time intervals. The bottom panel shows a SDO/HMI magnetogram for Carrington rotation 2116. The orange circle (indicated by SF) indicates the position N08E156 were presumably the parent eruption took place (see discussion below). The green, red and blue circles indicate the locations of Earth, STA and STB, respectively. The green, red and blue lines indicate the field lines connecting each spacecraft location with the solar surface as obtained by MAS/CORHEL. The yellow hatched areas indicate the longitude and latitude span of the CME at two different times. Several papers have been published discussing the observed properties of CME and SEP electrons and protons \citep{park2013, price2014, gomez2015}. \citet{zhao2018} modeled the behavior of SEPs released into interplanetary space. So we just briefly lay out those observed properties relevant to the modeling efforts contained in this paper.

The condition of the sun leading to the 2011 November 3 SEP event is somewhat complicated. There were multiple active regions (ARs) on the sun during the Carrington Rotation 2116. Notably were AR\#11333 located at N10W85 in the Heliocentric Earth Equatorial (HEEQ) coordinates and AR\#11339 located at N20E62. These regions produced several solar flares on the same day. One X1.9 flare occurred at 20:16 UT. It was associated with AR\#11339, but it was radio-silent. According to \citet{chen2013}, it only produced a failed filament eruption that remained confined by surrounding magnetic arcades. After that, GOES-15 observed four C-class X-ray flares starting at 22:12 UT, 22:28 UT, 22:56 UT, and 23:05 UT, and one M-class flare starting at 23:27 UT. \citet{park2013} suggested that any of these solar flares could separately contribute to the SEPs observed at the three spacecraft when they are magnetically connected.

There was a solar flare on the back side of the sun around the same time. GOES in the Earth orbit could not see it in X-ray, and STEREO does not have an X-ray instrument. \citet{nitta2013} used the 195 ${\rm \AA}$ flux obtained by the EUVI instrument on \textit{STB} as a proxy of X-ray emission from the flare. The X-ray flare was estimated to be located at N08E156 in the HEEQ coordinates (indicated by the yellow dot in Figure 1). The flare seemed to occur at a location not associated with any named AR, but one should note that the magnetic field measurements of the area shown in Figure 1 are already over two weeks old. The solar flare started around 22:11 UT and peaked at 22:41 UT with an estimated intensity between the levels of M4.7 and X1.4 class. It triggered a halo CME, which drove a shock as it was evident in Type II and Type III radio emissions observed by the WAVES instrument on \textit{Wind} and SWAVES on \textit{STA} and \textit{STB} \citep{gomez2015}. An EIT wave was observed propagating from the solar flare AR region, surpassing the magnetic footpoint of \textit{STA} by 22:21 UT and reaching the magnetic foot-point of \textit{STB} sometime later. Still, it seemed never to reach the magnetic foot-point of Earth. 

\begin{figure}
        \epsscale{0.5}
	\plotone{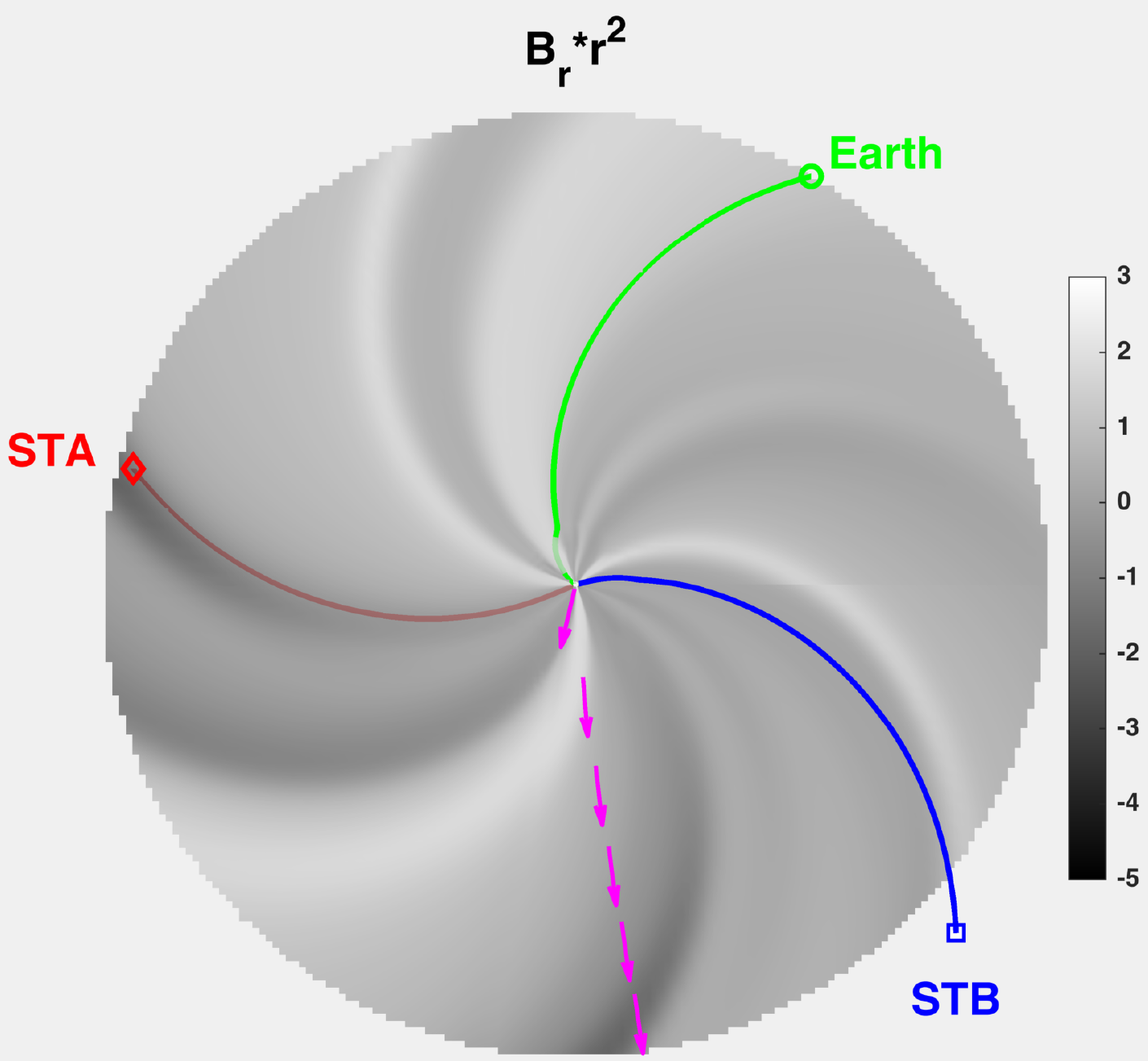}
	\epsscale{1.0}
	\plotone{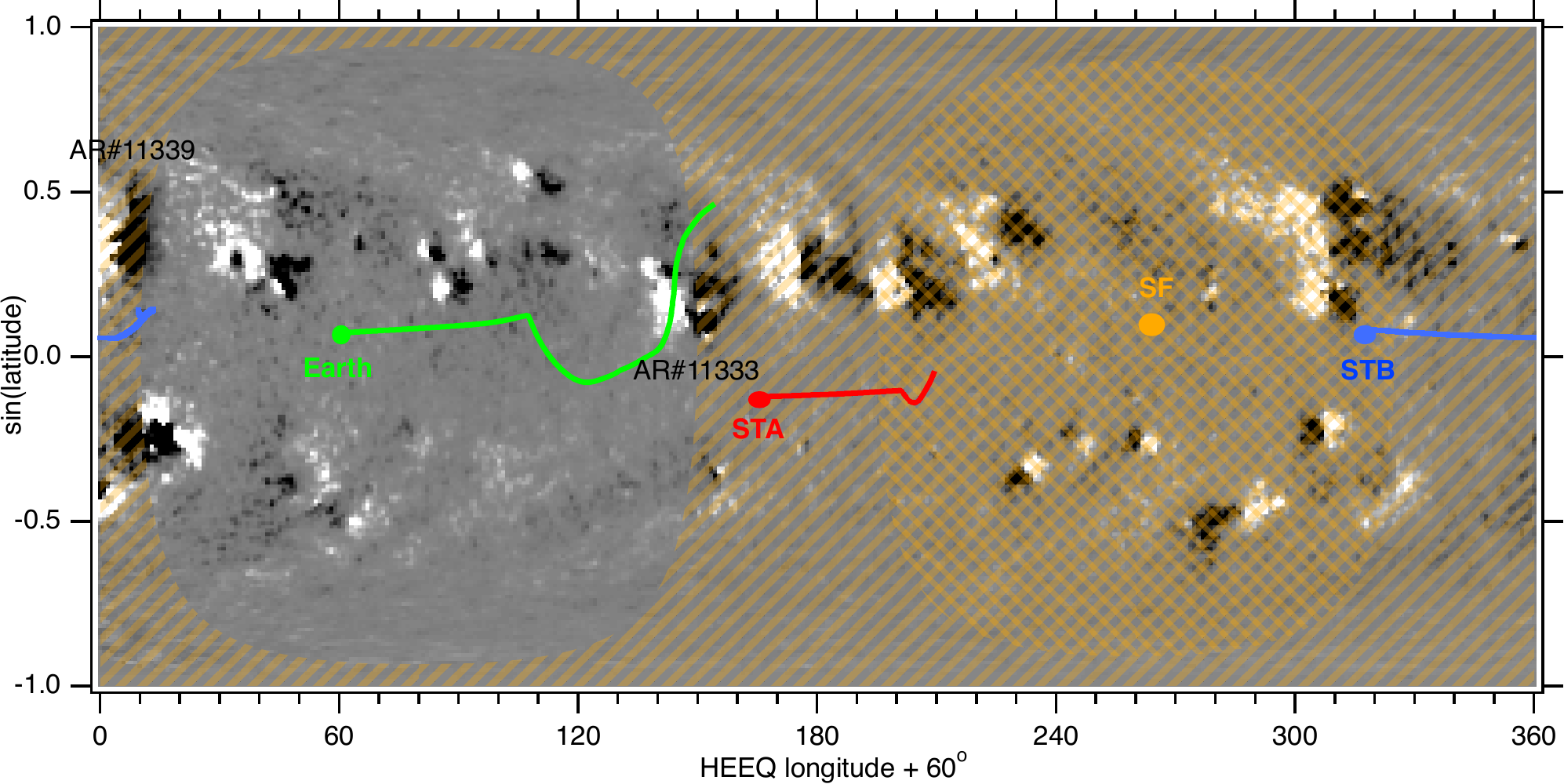}
	\caption{(Top) Equatorial view and (Bottom) projection on the solar surface of \textit{Earth}, \textit{STEREO-A} (STA), and \textit{STEREO-B} (STB) locations on 2011 November 3 (DOY 307) superposed on a background image showing the distribution of the radial magnetic field. The purple arrows denote the moving direction of CME shock. The curves are magnetic field lines that connect to the three locations. The yellow hatched areas indicate the latitude-longitude coverage of the CME shock at the time 23:40 UT (single hatched) and after DOY 308 05:54 UT (double hatched). A shift ($60^{\circ}$) of the heliocentric Earth equatorial(HEEQ) coordinate system in longitude is used, with the Earth at $60^{\circ}$ in longitude.  
		\label{fig:location}}
\end{figure}

The CME and its shock were clearly seen in coronagraph images obtained by all three spacecraft, which makes it possible to get a quality reconstruction of their 3D geometry. Figure 2 shows the shock surface at a few selected time instances. The first image frame with an observed CME shock occurred at 22:24 UT and the last one at 23:54 UT, during which the shock expanded in solar latitude and longitude as well as in radial distance up to $\sim$10 R$_\odot$. Initially, only \textit{STA} was connected to the west flank of the CME shock by a magnetic field line. By the time around 23:00 UT, \textit{STB} began to be connected, but the connection was brief, lasting roughly an hour. \textit{STB} was reconnected to the CME shock two days later on DOY 310 when the shock reached 1 AU as confirmed by in-situ plasma and magnetic field measurements on \textit{STB} \citep{gomez2015}. Earth never established magnetic connection with the shock, even though the magnetic field line appears to get under the yellow single-hatched region in Figure \ref{fig:location}, which indicates the maximum latitudinal-longitudinal coverage of the solar surface by the CME shock. This is because the maximum coverage by the CME shock occurred at a high altitude where the magnetic field line (in green color) is still slightly away from the shock. The footpoint of the magnetic field line connecting to Earth moves eastward significantly both in interplanetary space and in the solar corona, making it closer to the CME shock despite the large longitudinal difference between Earth and the solar flare. In contrast, the longitudinal motion of the magnetic field lines to \textit{STA} and \textit{STB} mainly occurs in interplanetary space. 

\begin{figure}
	\epsscale{1.0}
	\plottwo{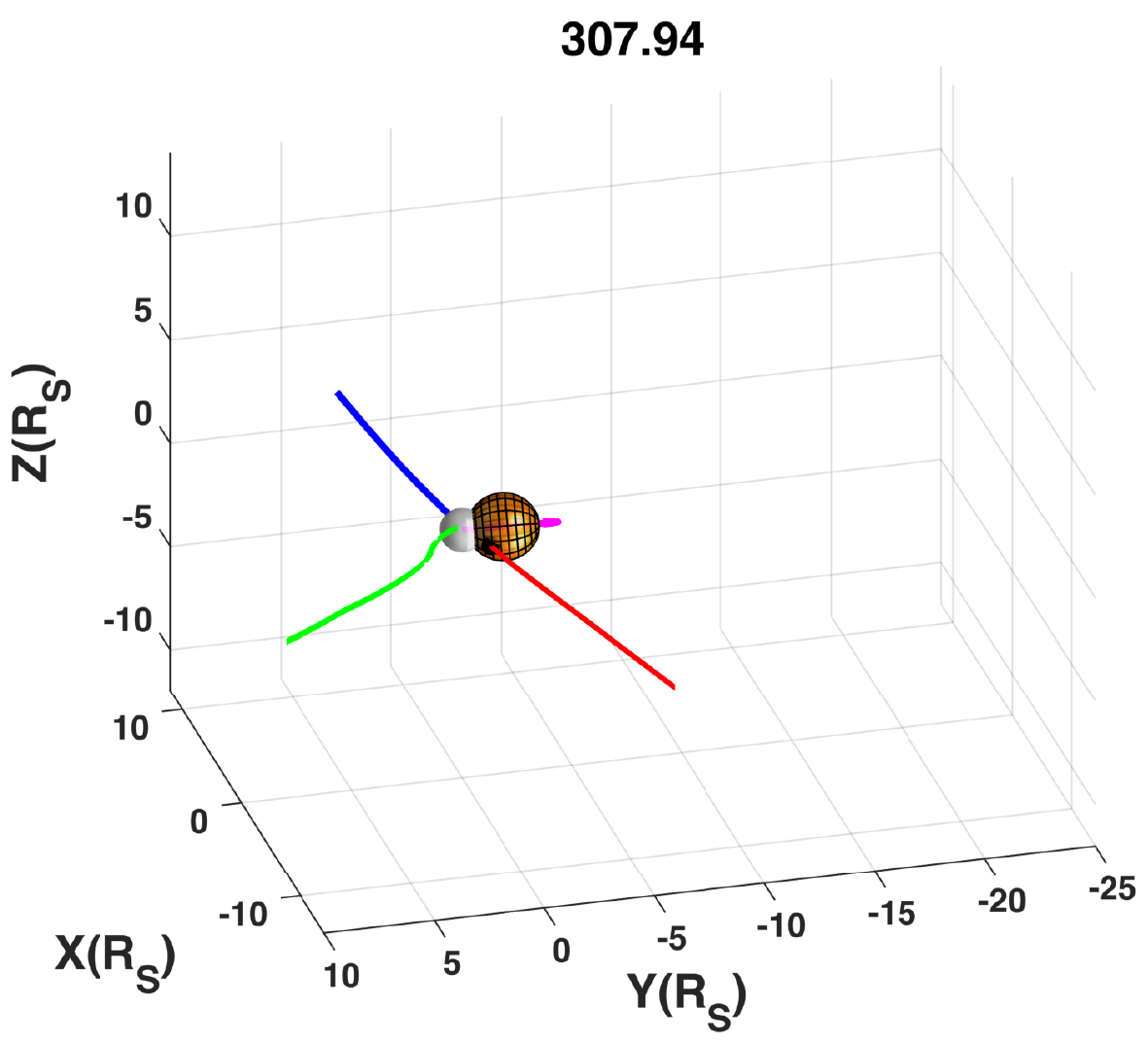}{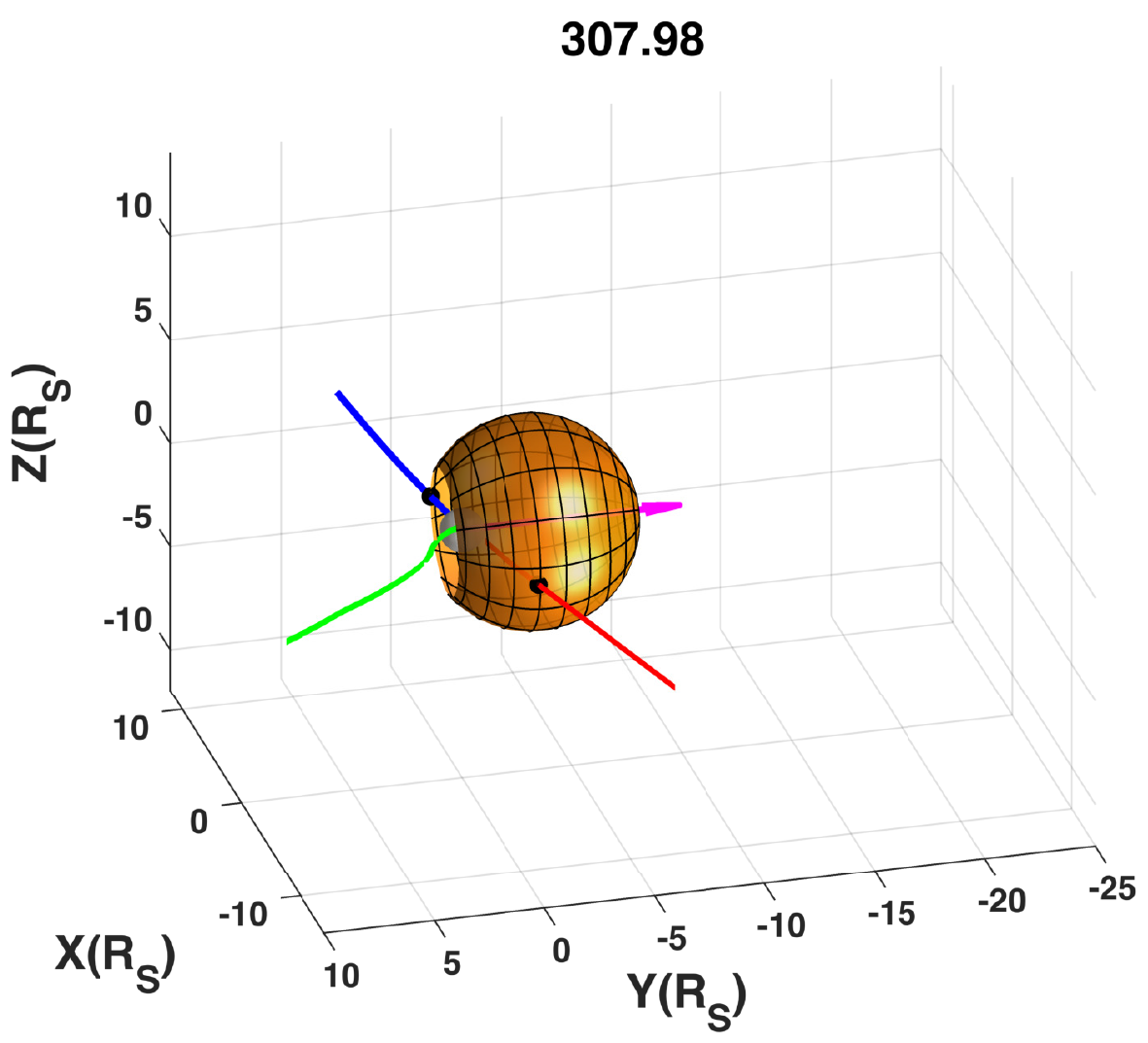}
	\plottwo{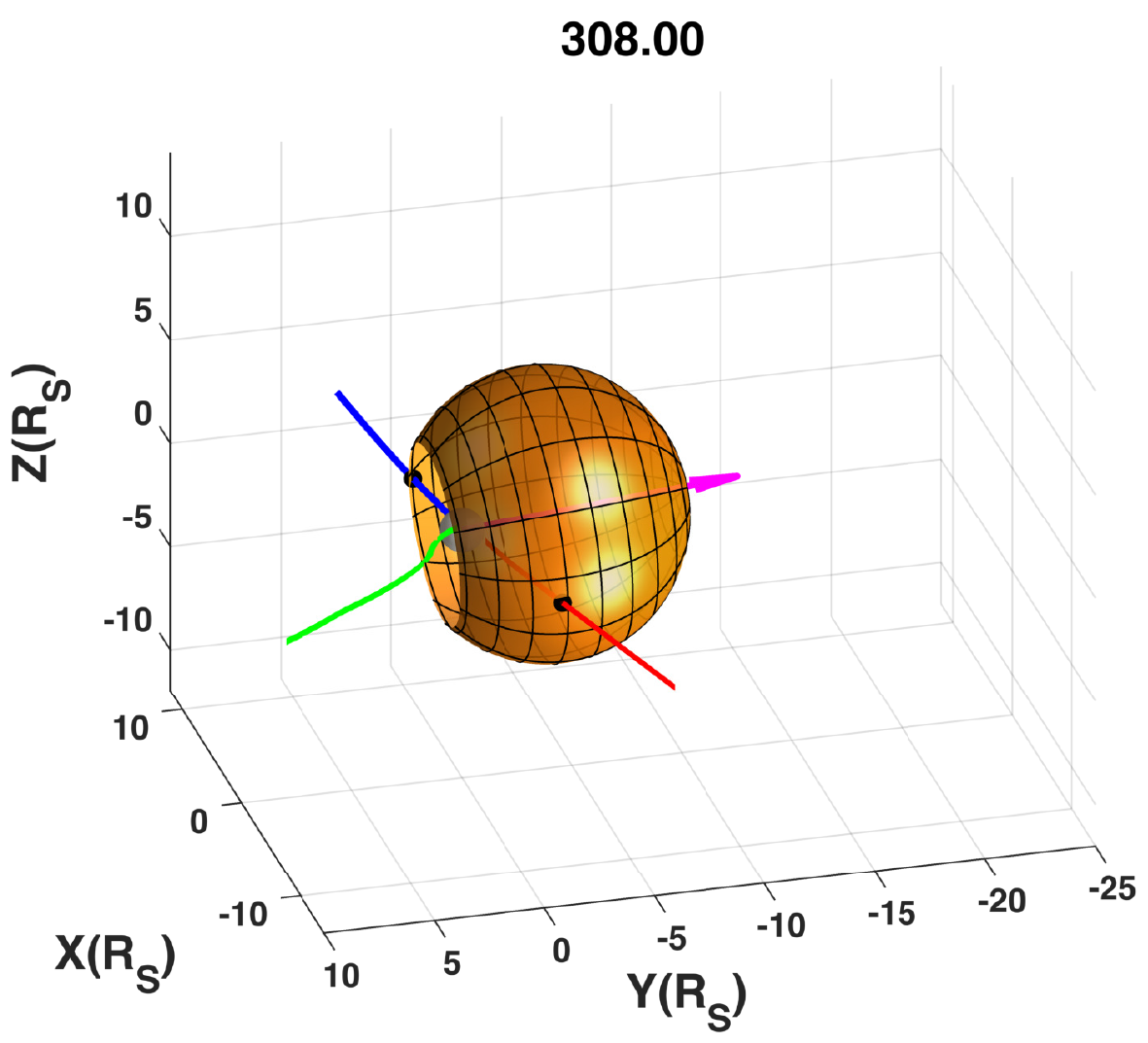}{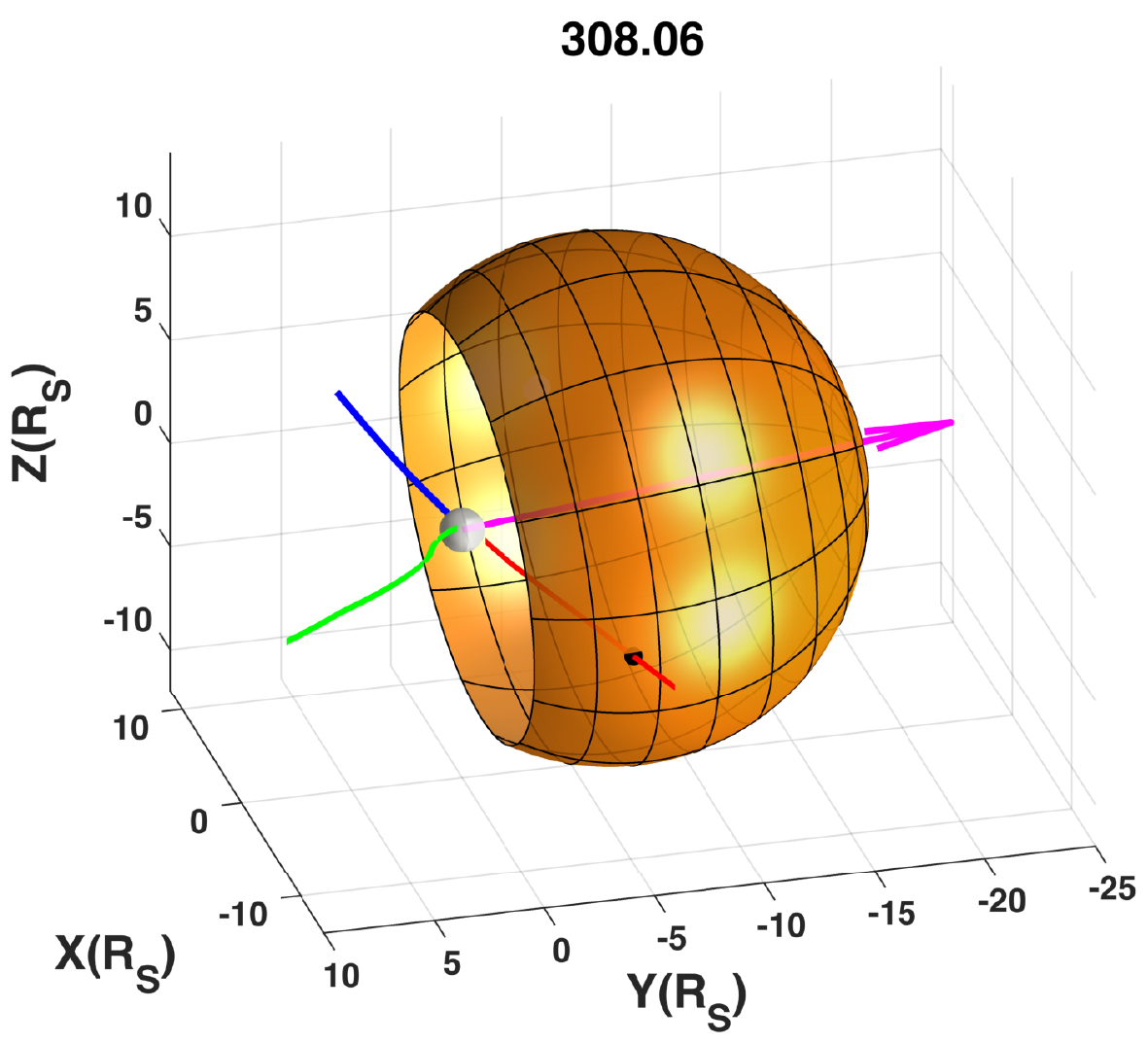}
	\plottwo{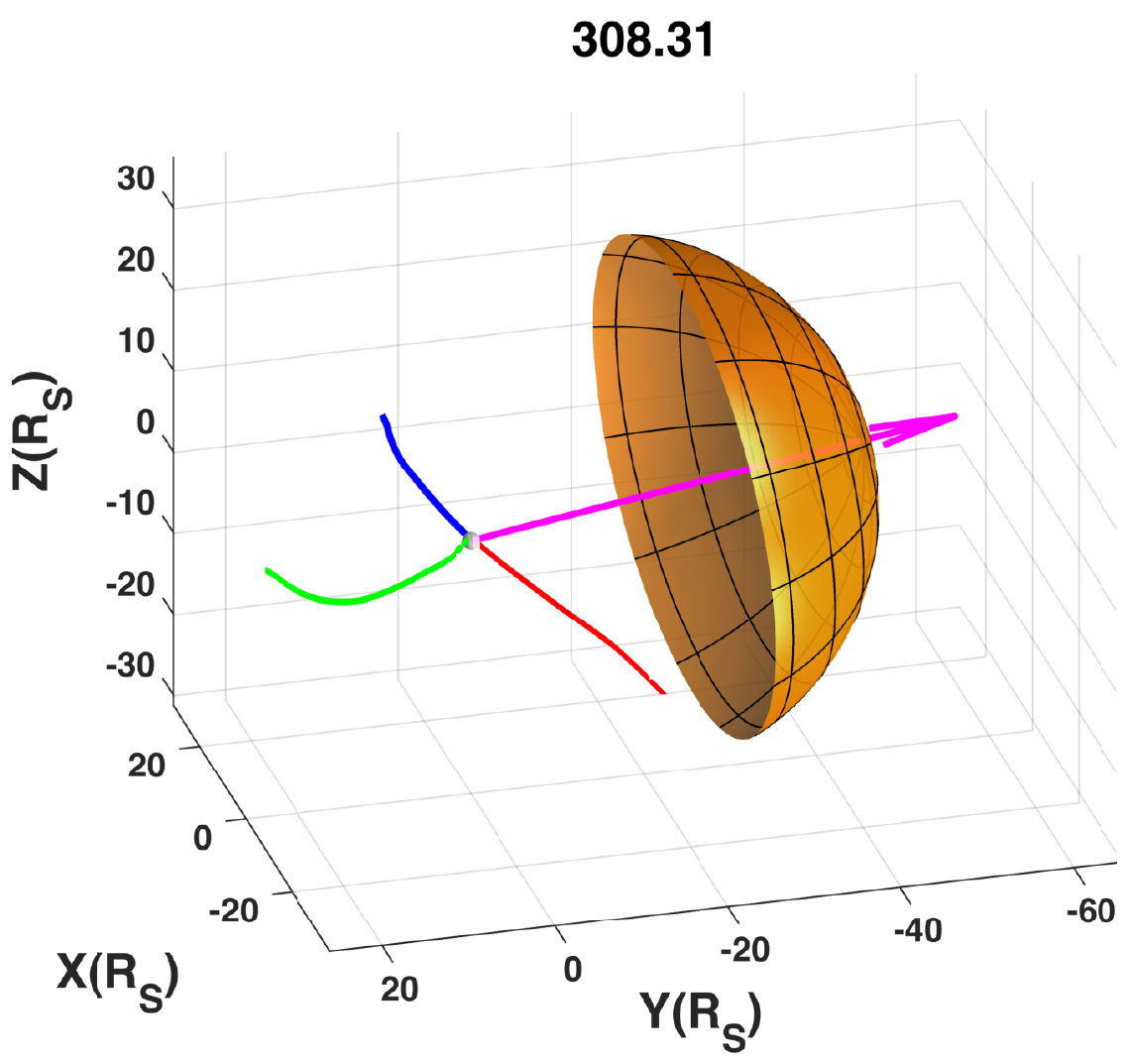}{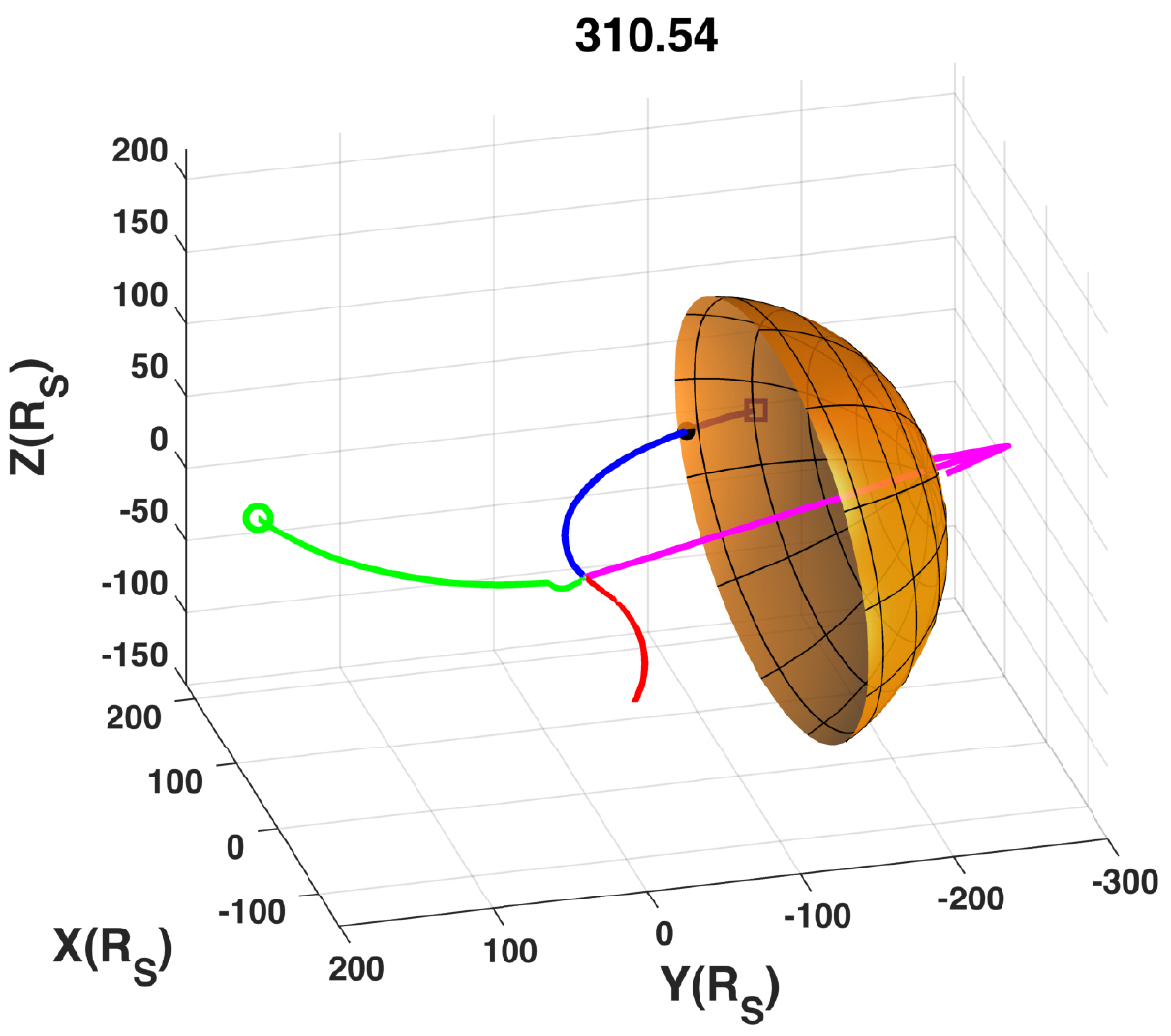}
	\caption{Ellipsoid CME shock surfaces (yellow color) at selected time intervals and their intersections (black dots) with magnetic field lines to Earth (green), STEREO-A (red), and STEREO-A (blue). Note that the scales for 308.31 and 310.54 differ from the rest. 
	\label{fig:surfaces}}
\end{figure}

\begin{figure}
	\epsscale{1.0}
	\plotone{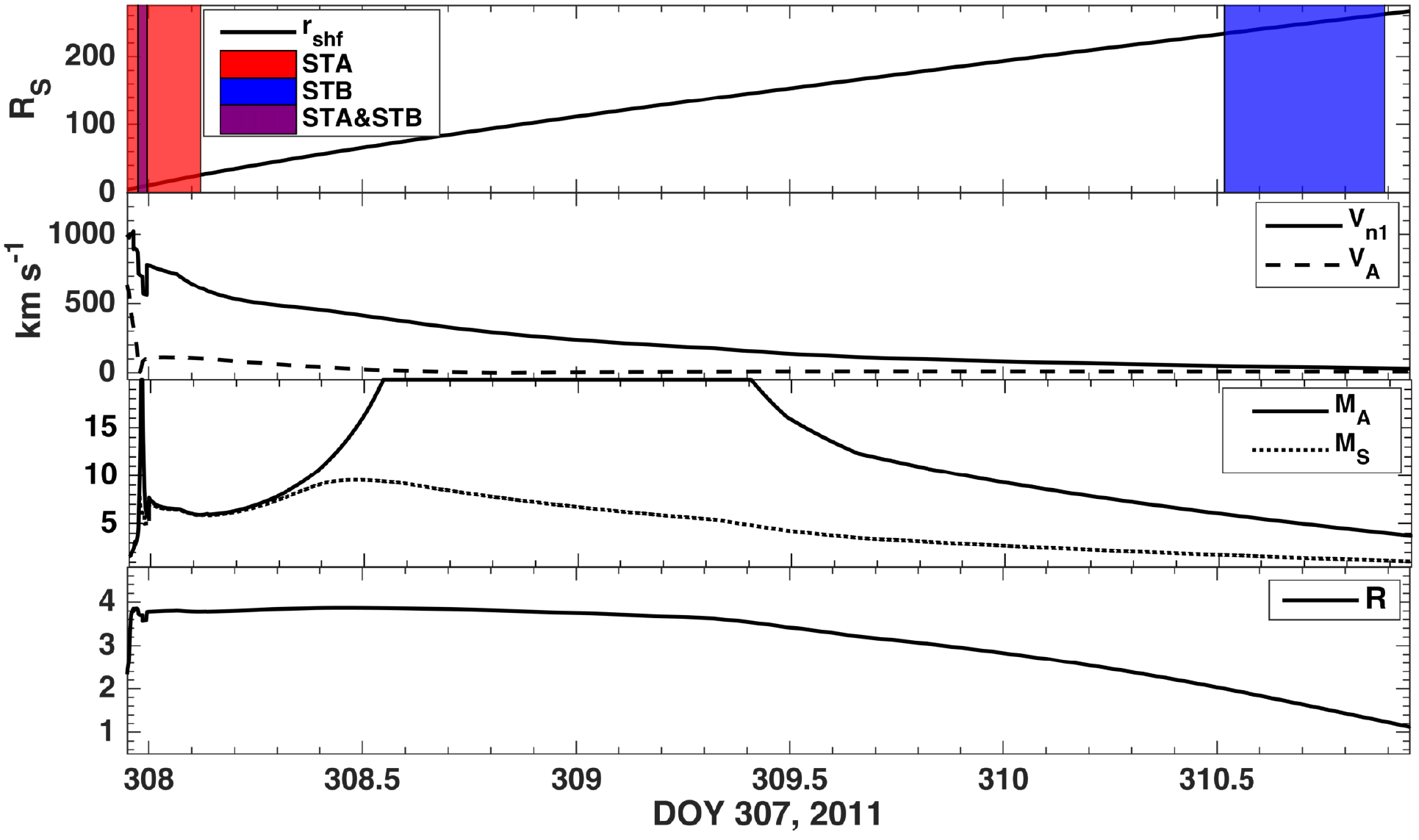}
	\caption{From top to bottom: Radial distance ($r_{shf}$), normal shock speed relative to upstream plasma ($V_{n1}$), upstream Alfv\'{e}n speed ($V_A$), Alfv\'{e}n Mach number ($M_A$), fast magnetosonic Mach number ($M_{S}$), and shock compression ratio ($R$) at the CME shock front as a function of time. The color shade regions in the top panel indicate the time when \textit{STA} and/or \textit{STB} are connected to the shock.
	\label{fig:shockfront}}
\end{figure}

\begin{table}[]
\caption{Key parameters regarding the CME shock propagation}
\begin{tabular}{|l|l|l|}
\hline
Solar flare rise time                                & $\Delta t_f$ & 5.0 min            \\
Initial CME speed                                    & $V_{cme0}$   & 1034 km s$^{-1}$   \\
Solar wind speed at 1 AU                             & $V_{1AU}$    & 370.0 km s$^{-1}$  \\
First critical time                                  & $\tau_{c1}$  & 90.3 min           \\
Radial distance of shock front at $\tau_{c1}$        & $r_{shf1}$   & 9.0 R$_\odot$          \\
Second critical time                                 & $\tau_{c2}$  & 203.6 min          \\
Radial distance of shock front at $\tau_{c2}$        & $r_{shf2}$   & 20.3 R$_\odot$         \\
Alfv\'{e}n speed in CME sheath at $\tau_{c2}$ & $V_{A2}$     & 80.0 km s$^{-1}$   \\
Sound speed in CME sheath $V_{S2}$ at $\tau_{c2}$  & $V_{S2}$     & 111.2 km s$^{-1}$  \\
Arrival time of shock front at 1 AU                          & $t_{1AU}$ & DOY 310 6:37 UT    \\
Speed of shock front at 1 AU                         & $V_{shf1AU}$ & 624.3 km s$^{-1}$ \\ \hline
\end{tabular}
\end{table}

According to the LASCO CME Catalog (http://cdaw.gsfc.nasa.gov/CME\_list/), the apparent plane-of-sky speed of the CME was 991 km s$^{-1}$. The DONKI catalog lists its speed as 1100 km/s. Our calculated speed using the time-evolving ellipsoid was 1034 km/s at the shock front at 22:24 UT when it was at 2.1 R$_\odot$ radial distance. With the input of its parent solar flare rise time of about 5 minutes, we could calculate the key parameters for the CME shock propagation as listed in Table 1. With the models for CME shock propagation and background plasma and magnetic field, we calculated the shock compression ratio at every point on the shock surface. Figure \ref{fig:shockfront} shows a few key parameters at the shock front along the direction the CME headed. The shock speed relative to the upstream solar wind $V_{n1}$ rapidly decreased with the time immediately after its initiation. At first, the decrease of $V_{n1}$ was mainly due to the acceleration of solar wind plasma in the corona. Once it reached the radial distance of $\sim20$ R$_\odot$ and after the second critical time, $\tau_{c2}$, the decrease of $V_{n1}$ came mainly from the slowdown of the CME. By the time it reached 1 AU, the shock speed was barely above the local Alfv\'en speed, and the shock compression became much weaker than it was at the beginning. 

\begin{figure}
	\includegraphics[height=2.9in]{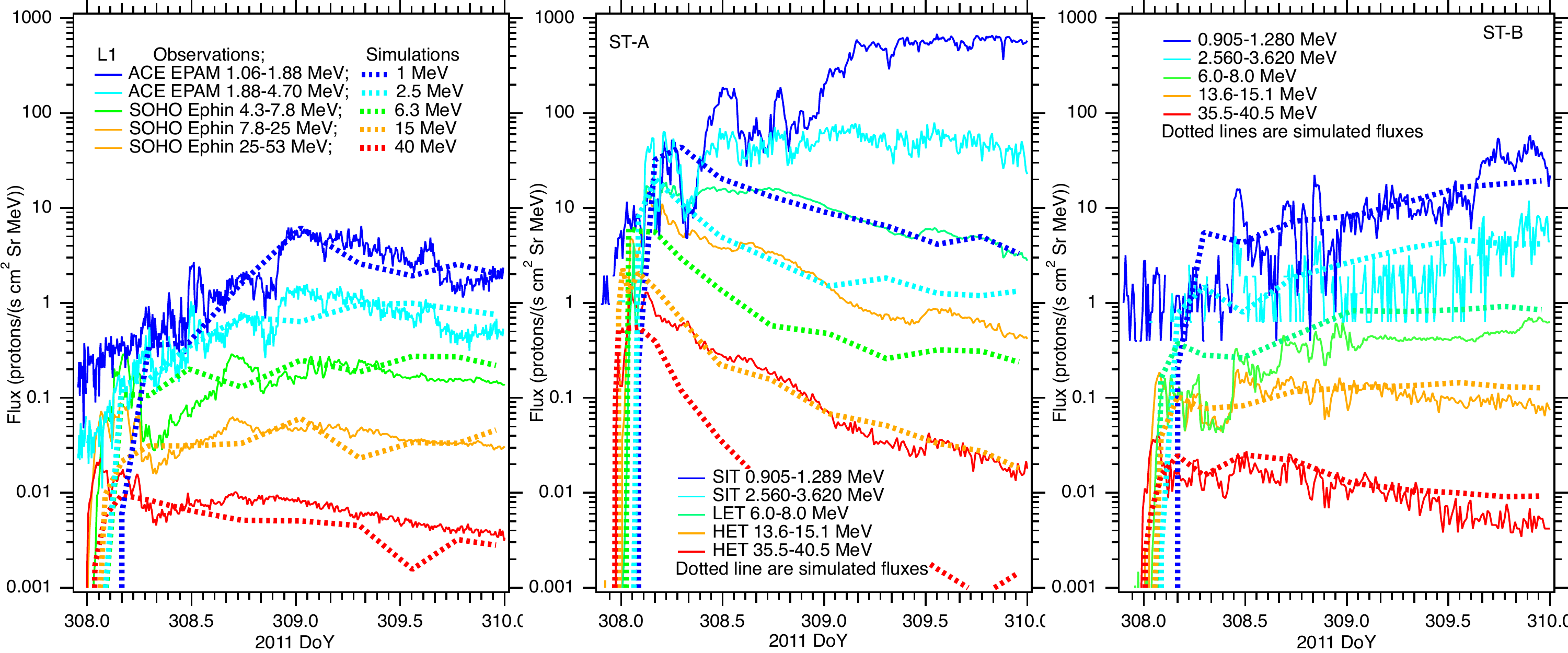}
	\caption{Proton fluxes at selected energy levels between $\sim$1 MeV to $\sim$50 MeV observed by \textit{SOHO}/\textit{ACE} at Earth's L1 point, \textit{STA} and \textit{STB} during the 2011 November 3 SEP event. The dotted traces are simulation results of absolute SEP fluxes for the three locations from one model run with a common set of model parameters. 
	\label{fig:sepobservations}}
\end{figure}

Figure \ref{fig:sepobservations} shows SEP proton fluxes observed at Earth's L1, \textit{STA} and \textit{STB} locations. A few selected energy channels from several instruments are plotted. The fluxes are direct measurements of particle intensity by the instrument. We do not show SEP electron data because our current code cannot adequately address the acceleration of electrons by shock waves. The behaviors of SEPs in this event have been reported in detail by \citet{gomez2015}. All the spacecraft registered enhancements of SEPs above $\sim$1 MeV to multiple tens MeV. Here we focus on the differences among observations at the three different locations. 

STEREO-A saw the highest intensities for all energies, except for $\sim$1 MeV proton intensities at \textit{STB} later on DOY 310, when the shock arrived at \textit{STB} locally \citep[see Figure 8 in][]{gomez2015}. At energies above $\sim$30 MeV, the intensities at all three locations rose to their peaks early and then gradually decayed afterward. This indicates that the high-energy particles are mainly produced in the corona, and little was produced in interplanetary space. Low-energy proton fluxes rose more gradually than high-energy protons, probably due to smaller particle mean free path at lower energies and a more continuous injection of particles from the shock. The low-energy proton fluxes peaked earlier at Earth than at \textit{STA}. The low-energy flux at \textit{STB} kept rising until the end of the graph because the shock was still approaching \textit{STB} at that time. This behavior indicates that low-energy particles can be produced in interplanetary space.

The enhancement of proton fluxes at \textit{STA} occurred the earliest compared to those at Earth and \textit{STB}. This, together with the higher particle fluxes at \textit{STA}, is consistent with the better and earlier magnetic connection established between \textit{STA} and the CME shock. \citet{gomez2015} made a detailed analysis of the SEP onset times. The onset time of SEPs at all three locations is linearly proportional to the reciprocal of particle speed. The velocity dispersion yields that the equivalent path length of the interplanetary magnetic field to the SEP source is nearly the same, around 1.9 AU, which is somewhat longer than the Parker spiral. The particle release time at the sun was also derived from the velocity dispersion. The particles arriving at \textit{STA} were first released from the sun at 22:21 UT, which is about 30 min earlier than those arriving at \textit{STB} and Earth L1 point. The onset of SEPs at Earth is the latest, consistent with the magnetic field footpoint to Earth being the farthest from the CME. 

\citet{gomez2015} reported a strong anti-sunward anisotropy of SEP electron fluxes during the rising phases of the event at all three locations, consistent with particles being released near the sun and propagating to 1 AU mainly along the magnetic field lines during the early phase of the event. The anisotropy of SEP protons from this event has not been analyzed.

\subsection{Simulation results}

\begin{figure}
	\epsscale{1.2}
	\plotone{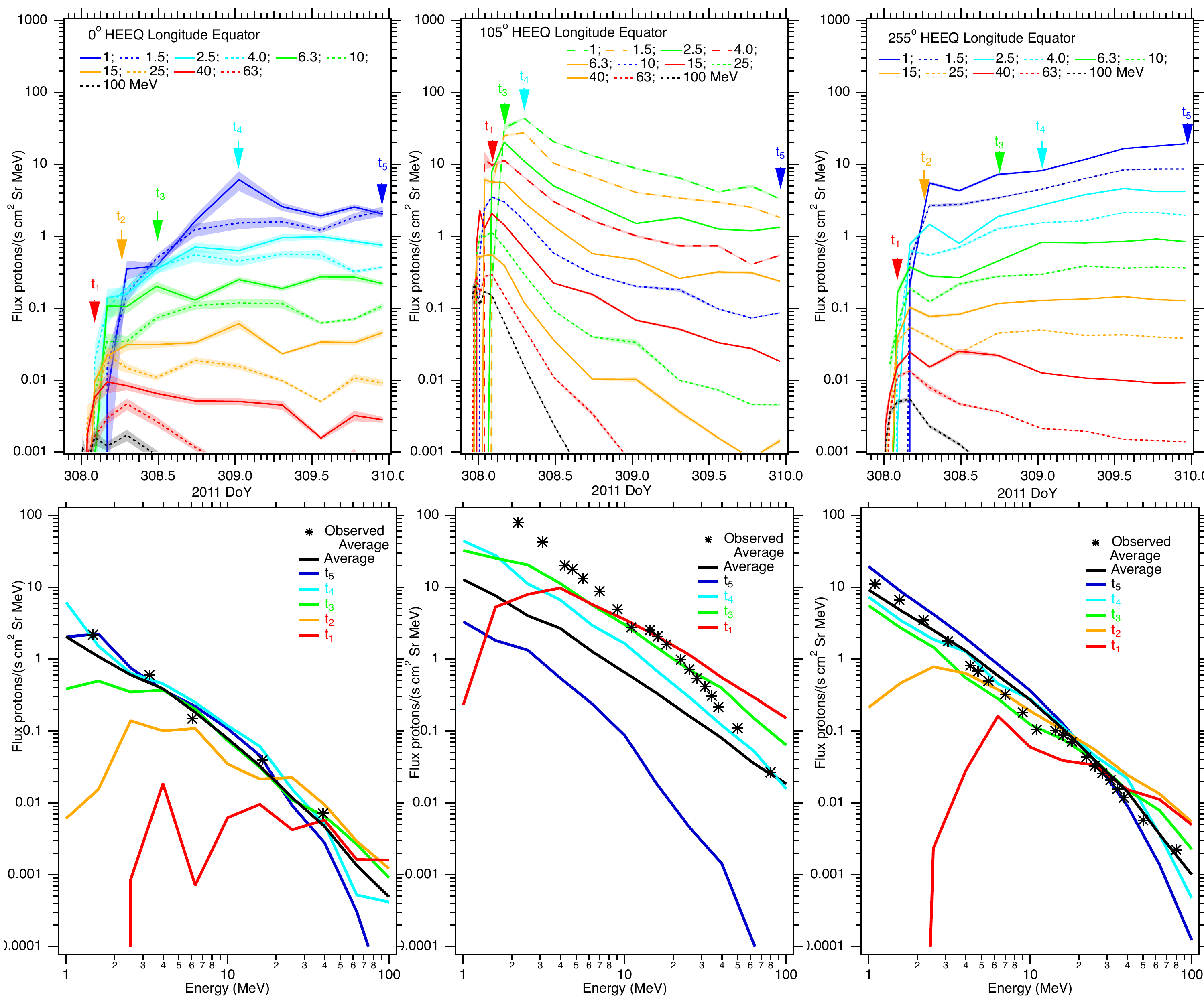}
	\caption{(Top row) Simulated time profiles of SEP proton flux expected at Earth (0$^\circ$), \textit{STA} (105$^\circ$) and \textit{STB} (255$^\circ$) for various energies as indicated in  the graph. The solid lines are color-coded to approximately match the observed time-intensity profiled in Figure \ref{fig:sepobservations} without consideration of energy. The arrows indicate the time of the energy spectra (Bottom row). The simulation assumes a constant radial particle mean free path of $\lambda_r = 200 R_S (p/1GV)^{1/3}$ and a $\alpha_\bot=0.37$ for perpendicular diffusion $\kappa_\bot$. 
	\label{fig:sima}}
\end{figure}

\begin{figure}
	\epsscale{1.0}
	\plottwo{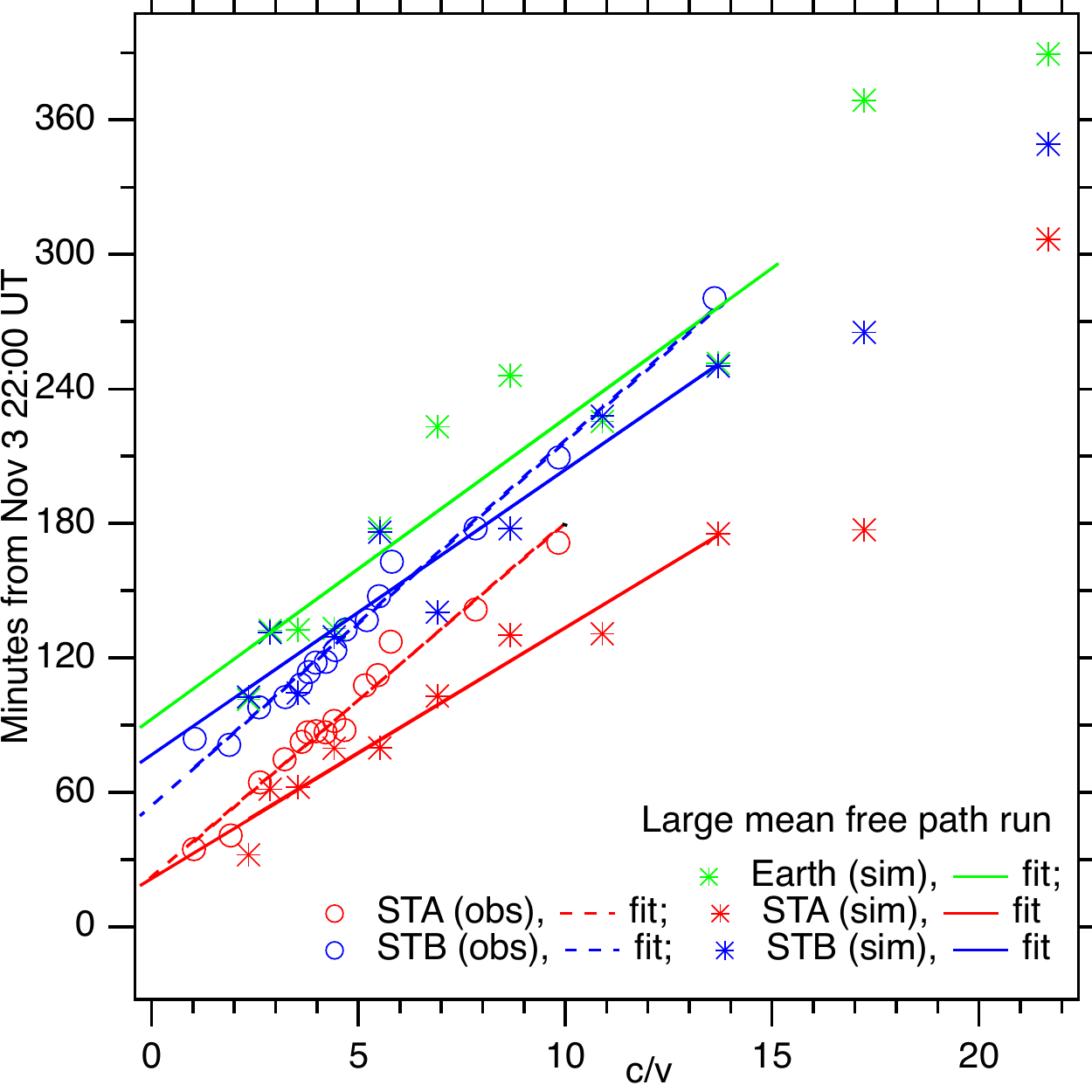}{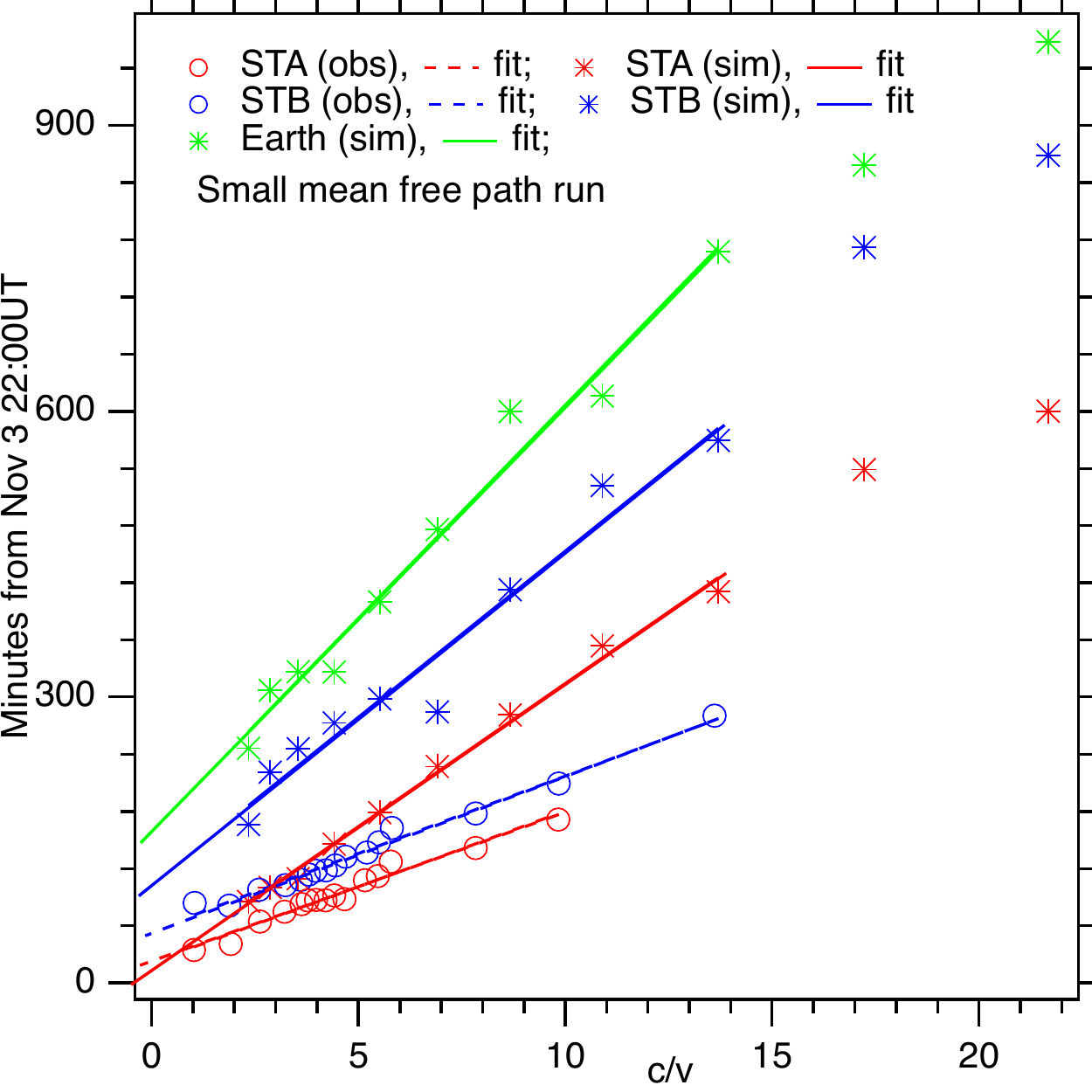}
	\caption{Onset time as a function of $c/v$ and linear fit. Data points for \textit{STA} and \textit{STB} observations are taken from \citet{gomez2015}. The left panel is based on a run with a large mean free path $\lambda_r = 200 R_S (p/1GV)^{1/3}$, and the right panel contains a run with a small mean free path $\lambda_r = 20 R_S (p/1GV)^{1/3}$.}
	\label{fig:onset}
\end{figure}

\begin{table}[]
\caption{Results of onset time velocity dispersion analysis} \label{onsetanalysis}
\begin{tabular}{llcc}
\hline
\hline
                                     & Location                      &      Apparent length (AU)        & Release Time UT)           \\ \hline
Observation                          & Earth                         & $1.86 \pm 1.44$ &   22:55 $\pm$ 15 min      \\
                                     & \textit{STA} & $1.90 \pm 0.09$ & 22:21 $\pm$ 4 min          \\
                                     & \textit{STB} & $1.96 \pm 0.06$ & 22:55 $\pm$ 3 min          \\ \hline
Simulation with large mean free path & Earth                         & $1.60 \pm 0.24$ & 23:33 $\pm$ 18 min         \\
                                     & \textit{STA} & $1.34 \pm 0.10$ & 22:22 $\pm$ 7 min          \\
                                     & \textit{STB\
                                     } & $1.52 \pm 0.20$ & 23:17 $\pm$ 12 min         \\ \hline
Simulation with small mean free path & Earth                         & $5.37 \pm 0.31$ & 1 day + 00:40 $\pm$ 19 min \\
                                     &\textit{STA} & $3.60 \pm 0.10$ & 22:14 $\pm$ 6 min          \\
                                     & \textit{STB} & $4.19 \pm 0.34$ & 23:45 $\pm$ 21 min         \\ \hline
\end{tabular}
\end{table}

\begin{figure}
	\epsscale{1.2}
	\plotone{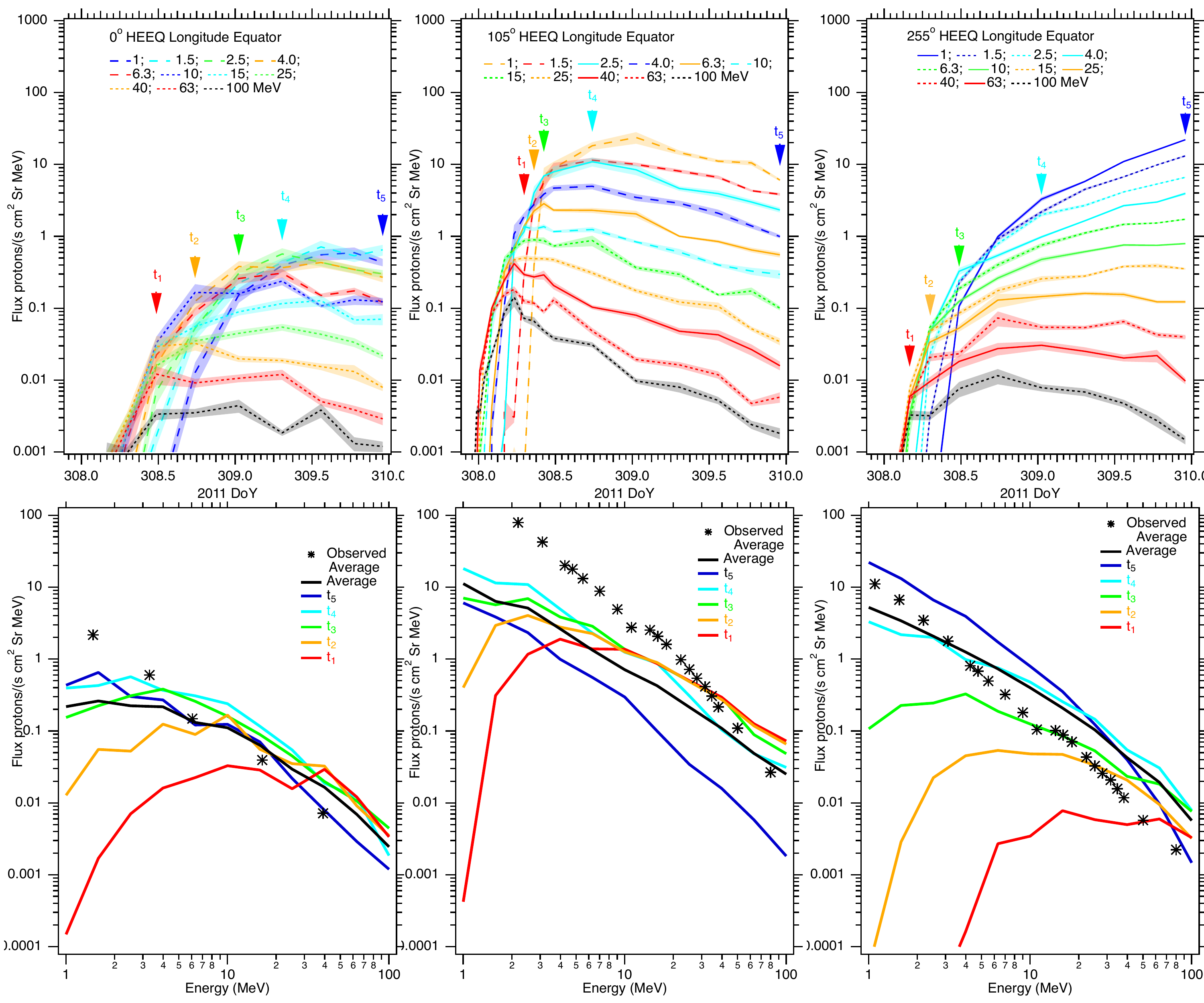}
	\caption{The same as Figure \ref{fig:sima} except for smaller radial mean free path of $\lambda_r = 20 R_S (p/1GV)^{1/3}$ and smaller $\alpha_\bot=0.074$ for perpendicular diffusion $\kappa_\bot$. 
	\label{fig:sim}}
\end{figure}

Figure \ref{fig:sima} shows the results of simulation of the time-intensity profiles expected at Earth, \textit{STA}, and \textit{STB} locations for various energies. Because the exact pitch angles of the measured particles are not known, we choose to only calculate the flux of particles outward along the magnetic field line. Unless the particle pitch angle distribution is beam-like, the omni-directional flux should not differ from the flux at 0 pitch angle by more than a factor of 2. We assign the  radial mean free path equal to 200 R$_\odot$ at 1 GV particle rigidity in the simulation, which is a relatively large mean free path. The mean free path scales as $p^{1/3}$ for other rigidities. Such a rigidity dependence comes from the quasi-linear pitch angle diffusion caused by the Kolmogorov turbulence spectrum. Compared with observations (see Figure \ref{fig:sepobservations}), the simulation results (the dotted traces) agree with observations at Earth and \textit{STB} quite well. The absolute intensity and time variation in the corresponding energy channels are consistent within a factor of $\sim$2. The bottom row of Figure \ref{fig:sima} shows the measured energy spectra averaged over the time interval shown in the figure (asterisks) and the simulated energy spectra at different times $t_{1} ... t_{5}$ indicated by the arrows (colored lines) and averaged over the whole period (black lines). The agreement between the calculated and observed energy spectra average over the time period is very good, as shown in the bottom left and right panels of Figure \ref{fig:sima}. This agreement is without any post-simulation normalization. Since the absolute particle intensity is sensitive to the assumption of particle injection, the match to observations without normalization suggests that our theory of shock acceleration from the thermal tail of post-shock solar wind plasma is reasonable, at least worth further testing.  

However, the calculation results for \textit{STA} do not reproduce the measured intensities well. If we try to pick a time-intensity profile to match the observed ones, we have to shift the energy. There are three pairs of time-intensity profiles that could be matched without consideration of energy level. For example, the calculation for 15 MeV protons could be compared to the observation in the 35.5-40.5 MeV channel on \textit{STA}. Other pairs could be matched with an energy level that differs by a factor of 2 to 3. Most mismatches to observations by \textit{STA} are in the low energy end. No calculated time-intensity profile can resemble what is seen in the $\sim$1 MeV proton flux by \textit{STA}. The calculated energy spectrum in the bottom middle panel of Figure \ref{fig:sima} looks completely different from the observed one, even in the spectral slope. The shift of energy and the difference in spectral slope to match the observations suggest that the particle mean free path and its energy dependence must be modified.  

Figure \ref{fig:onset} plots the particle onset time as a function of velocity reciprocal $c/v$. The onset time from our simulation is defined as the time when the calculated particle intensity rises above 1\% of its peak value. The observed onset time is taken from \citet{gomez2015}. Due to the presence of background in the measurements, the observed onset time was obtained using a different criterion. So the comparison can only be done qualitatively. On the left is for the run presented in Figure \ref{fig:sima} with the large mean free path of $200 R_S (p/1GV)^{1/3}$. The onset times are comparable to those determined from observations \citep{gomez2015} except for the slope, mostly due to the low-energy points, where the effect due to the background intensity is typically more severe. The inferred magnetic field line length and particle release time are listed in Table \ref{onsetanalysis}. Generally the agreement between observations and simulation with this choice of particle mean free path is reasonably good. The simulation predicts a generally smaller field line length because the mean free paths at low energies are larger than they should be. 

To investigate what causes the mismatch to the observations at \textit{STA}, we made another set of runs using a smaller mean free path and perpendicular diffusion coefficient. The results are shown in Figure \ref{fig:sim}, where we reduce the transport coefficients nearly by a factor of 10. The reduced transport coefficient mainly delays the rise of SEP intensity, but the calculated peak intensity is still in rough agreement with the observations. It works in the right direction in terms of the rise of low-energy particles around 1 MeV for the \textit{STA} location, although more improvement is still needed for 1 MeV protons. However, the rise time of high-energy protons appears to be delayed too much compared to the observations by \textit{STA}. The mean free path on the field line to \textit{STA} should be more steeply reduced toward low energies than the $p^{1/3}$ dependence. The delay of high-energy particles is also apparent in the calculated particle intensities for Earth and \textit{STB}. It seems we cannot reduce particle transport coefficients for high-energy particles. With the reduced particle mean free paths, the calculated particle onset time can be dramatically delayed, as shown in the right-side plot of Figure \ref{fig:onset}. The estimated apparent length of the magnetic field line is also increased. This demonstrates how the onset depends on particle mean free path controlled by the condition of magnetic field turbulence at the time. For this event, the observations point to a relatively large mean free path, in agreement with the simulation analysis in \citet{gomez2015}. 

\begin{figure}
	\includegraphics[height=2.9in]{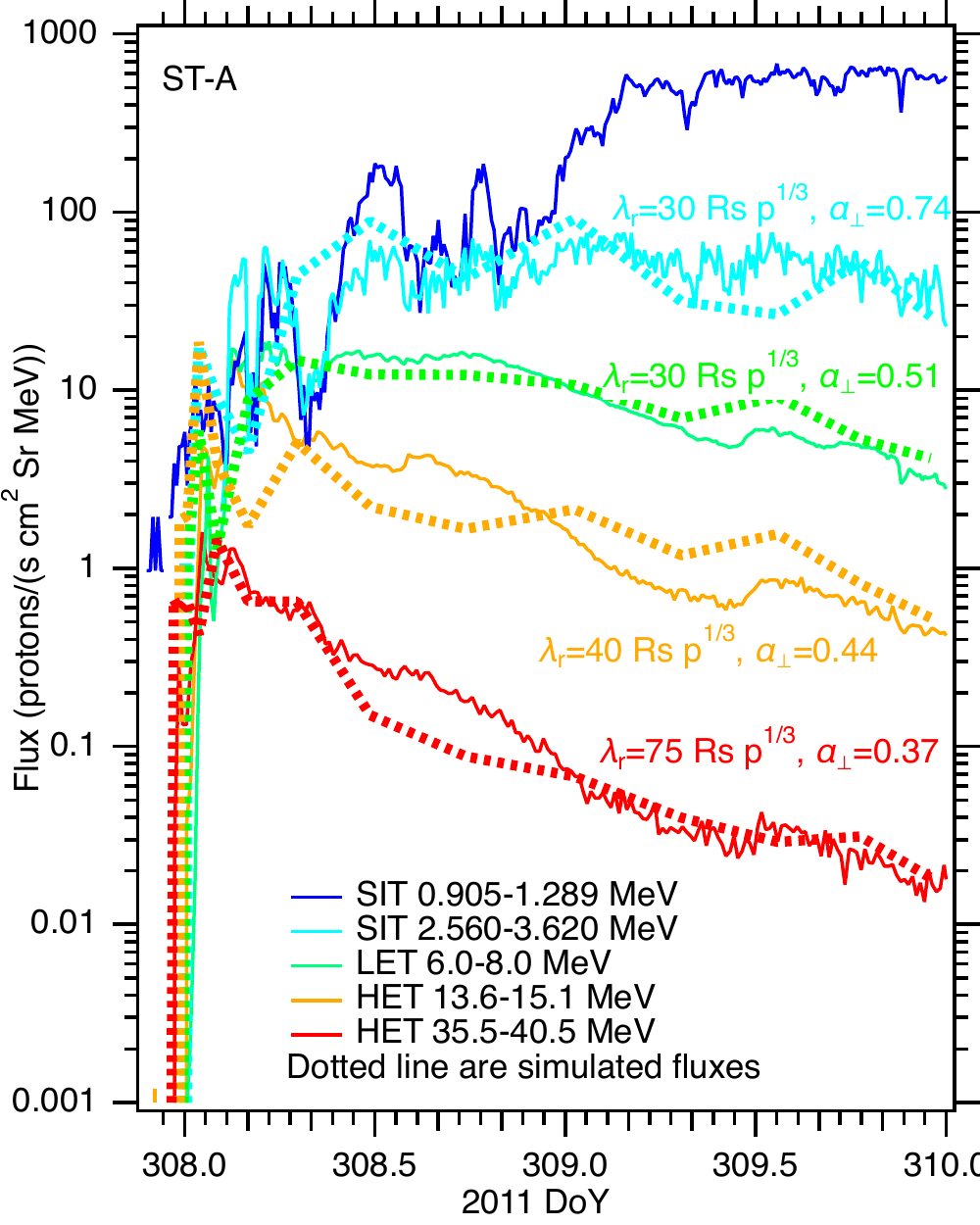}
	\caption{Proton fluxes at selected energy levels between $\sim$1 MeV to $\sim$50 MeV observed by \textit{STA}. The dotted traces are simulation results with mean free paths fine-tuned to match the observation individually in each energy channel. 
	\label{fluxsta1}}
\end{figure}

To fit the SEP intensity observations at \textit{STA}, we have made separate model runs for each individual energy channel, so that we can see what mean free path is needed for that particular energy at \textit{STA}. Figure \ref{fluxsta1} shows the time-intensity profile with a different mean free path. It appears that the parallel mean free path needs to be reduced while perpendicular diffusion needs to be enhanced at low energies. Still, the time-intensity profile of 0.905-1.289 MeV protons cannot be reproduced by any of our model runs. To explain why particles at \textit{STA} should have a smaller mean free path, we notice that \textit{STA} is well connected to the CME shock region, where the accelerated SEP density could be high enough to generate turbulence ahead of the shock. Most of the turbulence is produced by low-energy particles because they have high density. The low-energy particles resonantly interact with the self-generated turbulence, thus reducing their mean free path. High-energy particles or all particles on field lines to Earth and \textit{STB} do not experience resonant interaction with self-generated turbulence. Their mean free path should not be reduced. Currently, the code input particle mean free path and perpendicular diffusion coefficient as free parameters, but the spatial dependence and to some extent the rigidity dependence of particle mean free path are fixed. It does not include self-consistent wave-particle interaction treatment to modify the behavior of particle transport coefficients in the shock vicinity. Such improvement will be left for future work.

\begin{figure}
	\epsscale{0.6}
	\plotone{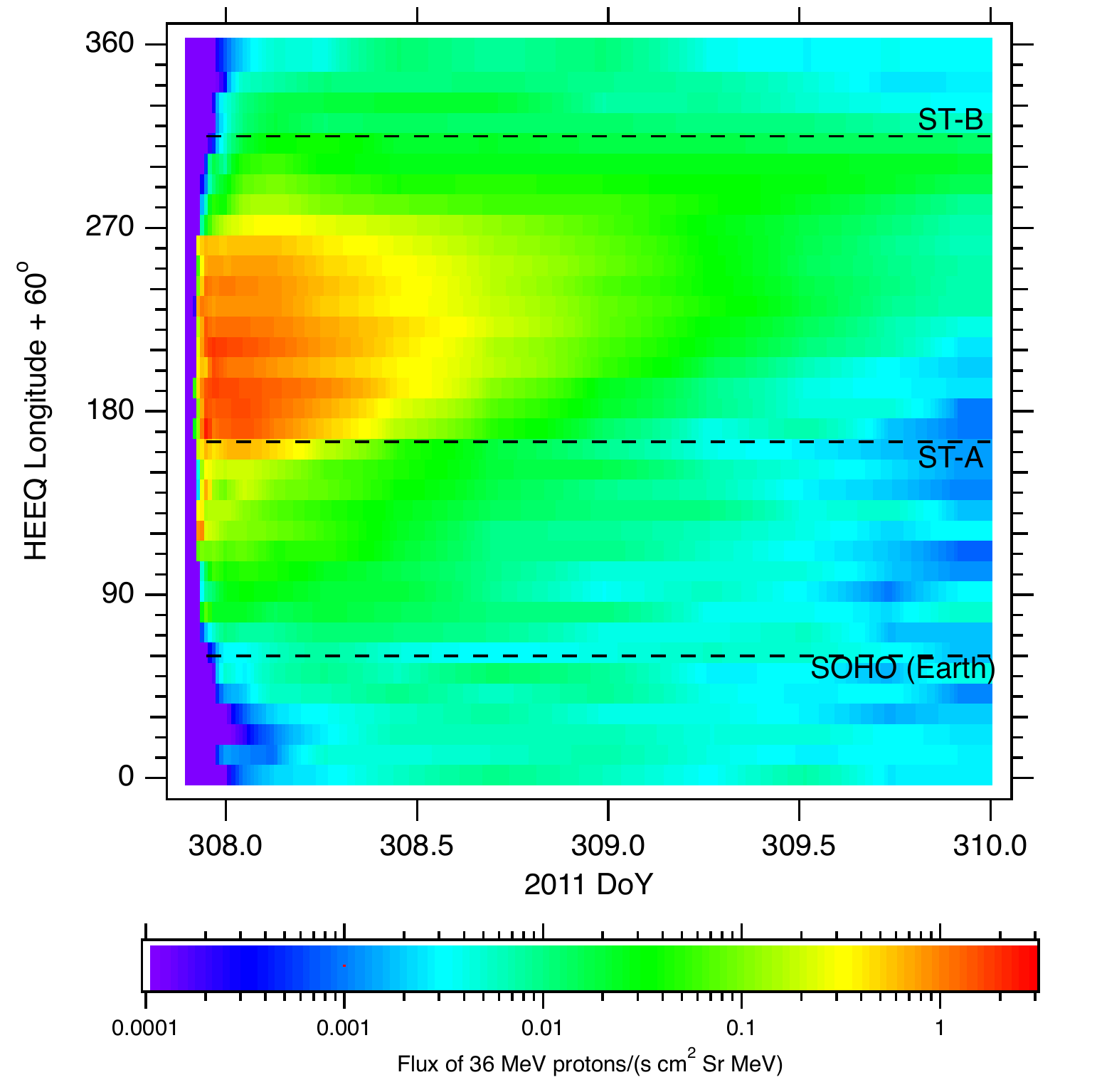}
	\caption{Calculated flux of 36 MeV protons as a function of time and longitude obtained from the simulation runs with a large mean free path $\lambda_r = 200 R_S (p/1GV)^{1/3}$ and a $\alpha_\bot=0.37$ for perpendicular diffusion coefficient $\kappa_\bot$.
		\label{fig:longtimea}}
\end{figure}

Figure \ref{fig:longtimea} shows the calculated flux of 36 MeV protons as a function of time and longitude in the solar equator and 1 AU for the 2011 November 3 event. The simulation is run with a large radial mean free path of $\lambda_r = 200 R_S (p/1GV)^{1/3}$ and a large perpendicular diffusion coefficient with $\alpha_\bot=0.37$. The plots exhibit how SEP intensity temporal variations would behave at various longitudes. The location of Earth, \textit{STA}, and \textit{STB} are indicated with the dashed lines. \textit{STA} was well connected to the CME shock at the beginning, and it sits near the core of SEP enhancement. \textit{STA} missed the region of most intensive SEP flux because the magnetic connection between \textit{STA} and CME shock did not last long, so the particles accelerated in the higher corona were not injected on the field line to \textit{STA}. Earth and \textit{STB} sit near the fringe area of the SEP core distribution because their magnetic field lines barely connected or missed the CME shock when it was low in the corona. If they had been 30$^\circ$ further out in longitude, the SEP onset time would have been much more delayed and the peak intensity much more reduced. Even so, the event is indeed circumsolar if a detection threshold is set below $\sim 0.005$ particles/(s cm$^2$ Sr MeV). The SEP flux at \textit{STB} did not decrease much with time because the CME shock was approaching within the two days. The shock crossed \textit{STB} afterward later on DOY 310.  

\begin{figure}
        \epsscale{0.6}
	\plotone{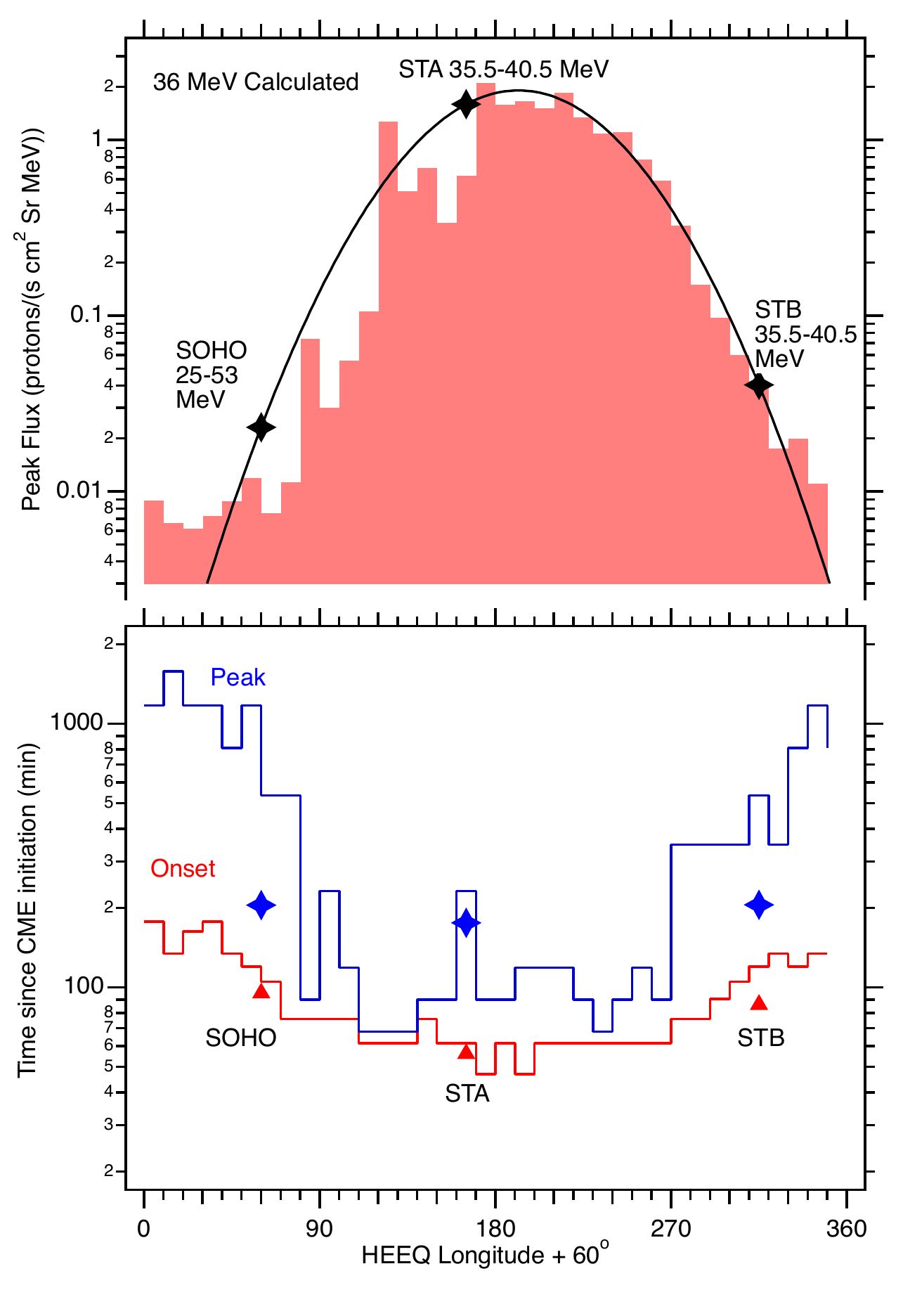}
	\caption{(Top) Peak flux of 36 MeV protons as a function of longitude and (Bottom) Peak flux time and onset time of 36 MeV protons as a function longitude together along with data from observations. The red curve at the top panel is a Gaussian distribution fitted to the observed peak fluxes. The results are obtained from the simulation runs with a large mean free path $\lambda_r = 200 R_S (p/1GV)^{1/3}$ and a $\alpha_\bot=0.37$ for perpendicular diffusion coefficient $\kappa_\bot$.
		\label{fig:peakfluxtimea}}
\end{figure}

The top panel of Figure \ref{fig:peakfluxtimea} shows the simulated peak flux of 36 MeV protons at 1 AU as a function of longitude. The observed peak fluxes in comparable energy channels on the spacecraft are plotted for comparison (black symbols). The agreement between simulation and measurements is reasonable good, although one should note that the energy of the data channel on \textit{SOHO} does not match those on the \textit{STA} and \textit{STB}. The onset time and peak flux time as a function of longitude are plotted in the bottom panel of Figure \ref{fig:peakfluxtimea}. The onset time agrees with the observations by the three spacecraft every well. The predicted peak flux times at Earth and \textit{STB} are somewhat delayed compared to the observed peak time. This difference is not a major concern because the peak time determination can carry a large uncertainty due to the broad time profiles around the peaks at those locations.

\begin{figure}
	\epsscale{0.6}
	\plotone{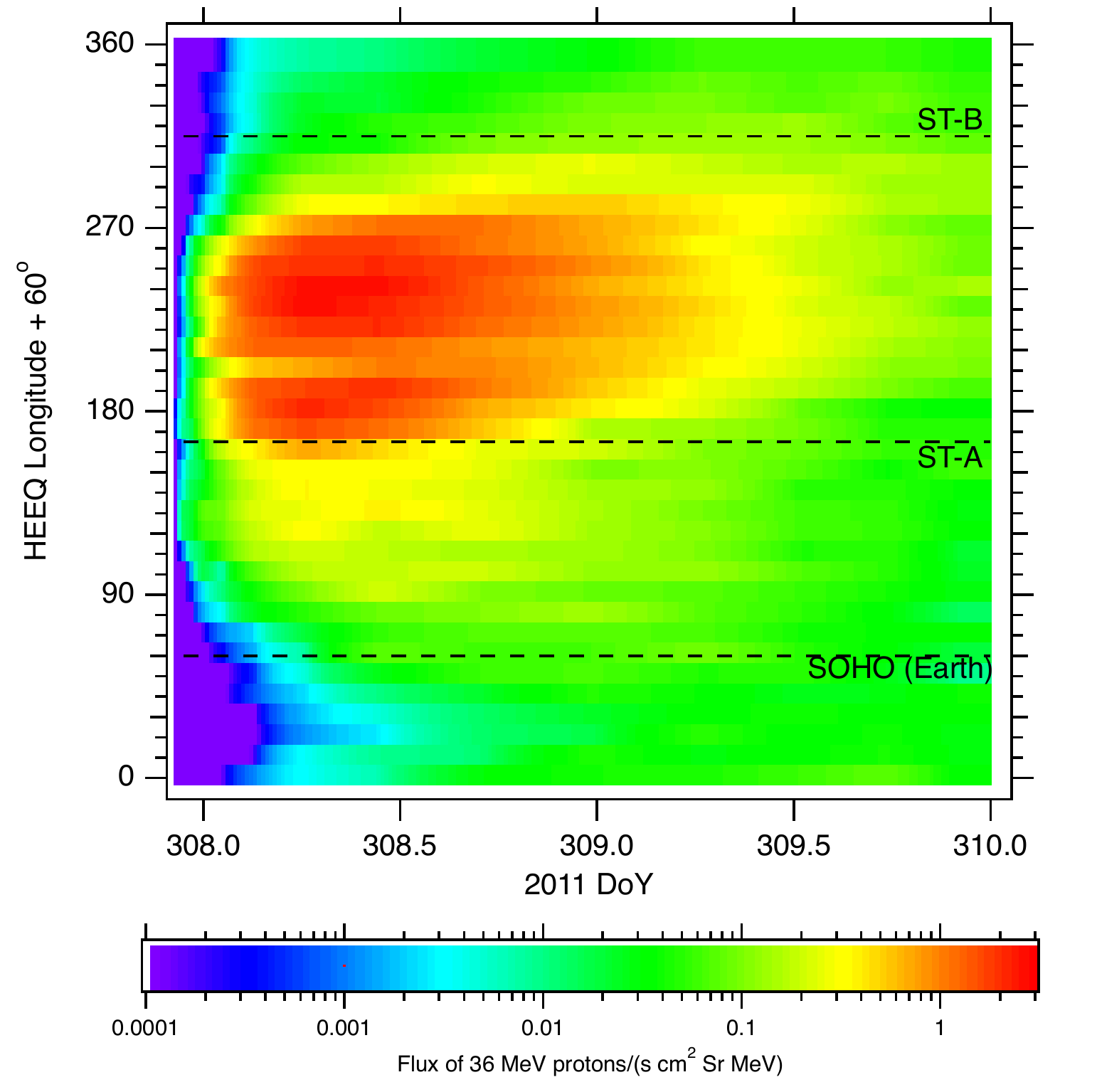}
	\caption{Calculated flux of 36 MeV protons as a function of time and longitude obtained from the simulation runs with a small mean free path $\lambda_r = 20 R_S (p/1GV)^{1/3}$ and a $\alpha_\bot=0.074$ for perpendicular diffusion coefficient $\kappa_\bot$.
		\label{fig:longtime}}
\end{figure}

\begin{figure}
        \epsscale{0.6}
	\plotone{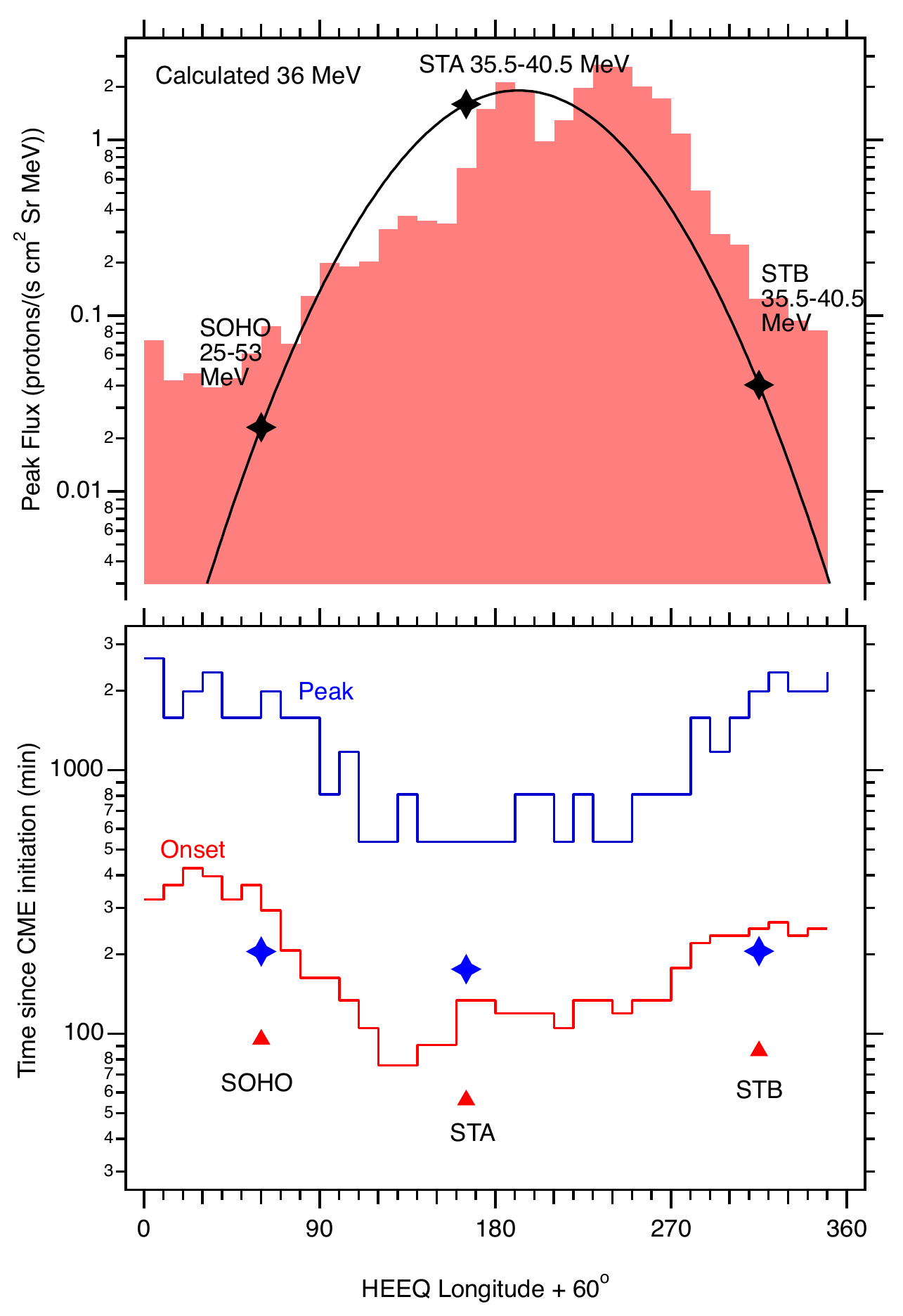}
	\caption{(Top) Peak flux of 36 MeV protons as a function of longitude and (Bottom) Peak flux time and onset time of 36 MeV protons as a function longitude together along with data from observations. The results are obtained from the simulation runs with a small mean free path $\lambda_r = 20 R_S (p/1GV)^{1/3}$ and a $\alpha_\bot=0.074$ for perpendicular diffusion coefficient $\kappa_\bot$. 
		\label{fig:peakfluxtime}}
\end{figure}

If the particle mean free paths are reduced, the SEP flux behavior would be different as shown in Figure \ref{fig:longtime} and Figure \ref{fig:peakfluxtime}. The onset time and peak flux time are very much delayed compared to the observations, suggesting it is not the case for the 2011 November 3 event. The peak flux in the core region does not change much, but the peak flux away from the core region is more elevated compared to the runs with a large mean free path. The elevated SEP flux lingers longer at all longitudes. It appears that the flux becomes nearly uniform across all longitudes after DOY 309.5, a sign of SEP reservoir formation \citep[][and references there in]{reames2013, qin2013}. Even though the perpendicular diffusion is also reduced in this run, with the reduced mean free path along the magnetic field, particles are held in the inner heliosphere for longer periods of time, causing the particles to diffuse across magnetic field lines while keeping the entire space with the elevated SEP intensity. 

\section{Summary and Discussion}\label{sec:disc}

This paper presents a new physics-based model for SEP production and propagation in a data-driven plasma and magnetic field structure of the solar corona and heliosphere with a data-driven propagating CME shock. The model rigorously solves the focused transport equation of energetic particles accelerated by CME shocks propagating through the corona and interplanetary medium. We can efficiently run the code with moderate computational power and solve the time-dependent 5-d phase space transport equation that includes particle perpendicular diffusion and acceleration by propagating shock waves with several advanced stochastic simulation techniques. The code can be used to predict the SEP intensity for any particle energy and pitch angle at any location in the heliosphere.  

The code is first applied to the circumsolar SEP event on 2011 November 3, which was observed \textit{STA}, \textit{STB} and near-Earth spacecraft located at widely separated heliographic longitudes saw SEP enhancements within an hour nearly simultaneously. The code takes the input of corona and heliospheric plasma and magnetic field from the MAS/CORHEL MHD model driven by solar photospheric magnetic field measurements with an observed CME shock determined from coronagraph images. With an assumption of particle injection from post-shock heated thermal tail solar wind ions, the predicted time-intensity profiles can fit the SEP observations from three spacecraft locations close to $\sim$1 AU in near-the-ecliptic orbits. It demonstrates that SEPs seen at widely separated longitudes are produced by a single CME shock going through the solar corona and interplanetary medium. Magnetic field configuration in the corona and heliosphere and the size of CME shock initiate an extensive coverage of heliographic longitude, but it still needs perpendicular diffusion to spread the particles to unconnected magnetic field lines. We need to set proper magnitudes of particle pitch angle diffusion and perpendicular diffusion across magnetic fields to reproduce the observed SEPs. Based on the preliminary runs, we found that the code may be able to predict the absolute peak intensity of SEP within the same order of magnitude without renormalization.


\begin{acknowledgments}
This work was partially supported by NASA Grants 80NSSC21K0004, 80NSSC20K0098,  80NSSC20K0086, 80NSSC19K1254.  JZ was partially supported under the auspices of the U.S. Department of Energy by Lawrence Livermore National Laboratory under contract DE-AC52-07NA27344. PR gratefully acknowledges support from NASA (80NSSC20C0187, 80NSSC20K1285, 80NSSC20K1403, and 80NSSC22CA210) and NRL (N00173-17-C-2003). D.L. acknowledges support from NASA Living With a Star (LWS) programs NNH17ZDA001N-LWS and NNH19ZDA001N-LWS, the Goddard Space Flight Center Internal Scientist Funding Model (competitive work package) program, and the Heliophysics Innovation Fund (HIF) program. LB acknowledges support from NASA  80NSSC19K1235.
\end{acknowledgments}

%
%
%

\bibliography{ref}{}

\begin{thebibliography}{}
\expandafter\ifx\csname natexlab\endcsname\relax\def\natexlab#1{#1}\fi
\providecommand{\url}[1]{\href{#1}{#1}}
\providecommand{\dodoi}[1]{doi:~\href{http://doi.org/#1}{\nolinkurl{#1}}}
\providecommand{\doeprint}[1]{\href{http://ascl.net/#1}{\nolinkurl{http://ascl.net/#1}}}
\providecommand{\doarXiv}[1]{\href{https://arxiv.org/abs/#1}{\nolinkurl{https://arxiv.org/abs/#1}}}

\bibitem[{{Bieber} {et~al.}(1994){Bieber}, {Matthaeus}, {Smith}, {Wanner},
  {Kallenrode}, \& {Wibberenz}}]{Bieberetal1994}
{Bieber}, J.~W., {Matthaeus}, W.~H., {Smith}, C.~W., {et~al.} 1994, \apj, 420,
  294, \dodoi{10.1086/173559}

\bibitem[{{Bj{\"o}rk}(2015)}]{bjork2015}
{Bj{\"o}rk}, T. 2015, arXiv e-prints, arXiv:1512.08912.
\newblock \doarXiv{1512.08912}

\bibitem[{{Book}(1987)}]{book1987}
{Book}, D.~L. 1987, {NRL (Naval Research Laboratory) Plasma Formulary, revised}

\bibitem[{{Bothmer} \& {Schwenn}(1998)}]{bothmer1998}
{Bothmer}, V., \& {Schwenn}, R. 1998, Annales Geophysicae, 16, 1,
  \dodoi{10.1007/s00585-997-0001-x}

\bibitem[{{Caplan} {et~al.}(2017){Caplan}, {Miki{\'c}}, {Linker}, \&
  {Lionello}}]{caplan2017}
{Caplan}, R.~M., {Miki{\'c}}, Z., {Linker}, J.~A., \& {Lionello}, R. 2017, in
  Journal of Physics Conference Series, Vol. 837, Journal of Physics Conference
  Series, 012016, \dodoi{10.1088/1742-6596/837/1/012016}

\bibitem[{{Caprioli} \& {Spitkovsky}(2014)}]{caprioli2014}
{Caprioli}, D., \& {Spitkovsky}, A. 2014, \apj, 783, 91,
  \dodoi{10.1088/0004-637X/783/2/91}

\bibitem[{{Chen} {et~al.}(2013){Chen}, {Ma}, \& {Zhang}}]{chen2013}
{Chen}, H., {Ma}, S., \& {Zhang}, J. 2013, \apj, 778, 70,
  \dodoi{10.1088/0004-637X/778/1/70}

\bibitem[{{Corona-Romero} {et~al.}(2013){Corona-Romero}, {Gonzalez-Esparza}, \&
  {Aguilar-Rodriguez}}]{corona-romero2013}
{Corona-Romero}, P., {Gonzalez-Esparza}, J.~A., \& {Aguilar-Rodriguez}, E.
  2013, \solphys, 285, 391, \dodoi{10.1007/s11207-012-0103-9}

\bibitem[{{Corona-Romero} {et~al.}(2017){Corona-Romero}, {Gonzalez-Esparza},
  {Perez-Alanis}, {Aguilar-Rodriguez}, {de-la-Luz}, \&
  {Mejia-Ambriz}}]{corona-romero2017}
{Corona-Romero}, P., {Gonzalez-Esparza}, J.~A., {Perez-Alanis}, C.~A., {et~al.}
  2017, Space Weather, 15, 464, \dodoi{10.1002/2016SW001489}

\bibitem[{{Downs} {et~al.}(2016){Downs}, {Lionello}, {Miki{\'c}}, {Linker}, \&
  {Velli}}]{downs2016}
{Downs}, C., {Lionello}, R., {Miki{\'c}}, Z., {Linker}, J.~A., \& {Velli}, M.
  2016, \apj, 832, 180, \dodoi{10.3847/0004-637X/832/2/180}

\bibitem[{{Dr\"{o}ge}(1994)}]{Droge1994}
{Dr\"{o}ge}, W. 1994, \apjs, 90, 567, \dodoi{10.1086/191876}

\bibitem[{{Dr{\"o}ge} {et~al.}(2014){Dr{\"o}ge}, {Kartavykh}, {Dresing},
  {Heber}, \& {Klassen}}]{droge2014}
{Dr{\"o}ge}, W., {Kartavykh}, Y.~Y., {Dresing}, N., {Heber}, B., \& {Klassen},
  A. 2014, Journal of Geophysical Research (Space Physics), 119, 6074,
  \dodoi{10.1002/2014JA019933}

\bibitem[{{Dr{\"o}ge} {et~al.}(2010){Dr{\"o}ge}, {Kartavykh}, {Klecker}, \&
  {Kovaltsov}}]{Drogeetal2010}
{Dr{\"o}ge}, W., {Kartavykh}, Y.~Y., {Klecker}, B., \& {Kovaltsov}, G.~A. 2010,
  \apj, 709, 912, \dodoi{10.1088/0004-637X/709/2/912}

\bibitem[{{Drury}(1983)}]{Drury1983}
{Drury}, L.~O. 1983, Reports on Progress in Physics, 46, 973,
  \dodoi{10.1088/0034-4885/46/8/002}

\bibitem[{{Engelbrecht} {et~al.}(2022){Engelbrecht}, {Effenberger},
  {Florinski}, {Potgieter}, {Ruffolo}, {Chhiber}, {Usmanov}, {Rankin}, \&
  {Els}}]{engelbrecht2022}
{Engelbrecht}, N.~E., {Effenberger}, F., {Florinski}, V., {et~al.} 2022, \ssr,
  218, 33, \dodoi{10.1007/s11214-022-00896-1}

\bibitem[{{Farris} \& {Russell}(1994)}]{farris1994}
{Farris}, M.~H., \& {Russell}, C.~T. 1994, \jgr, 99, 17681,
  \dodoi{10.1029/94JA01020}

\bibitem[{{Forman} \& {Drury}(1983)}]{forman1983}
{Forman}, M.~A., \& {Drury}, L.~O. 1983, in International Cosmic Ray
  Conference, Vol.~2, International Cosmic Ray Conference, 267

\bibitem[{{Freidlin}(1985)}]{Freidlin1985}
{Freidlin}, M. 1985, {Functional Integration and Partial Differential Equation}
  (Princeton)

\bibitem[{{Gardiner}(1983)}]{Gardiner1983}
{Gardiner}, C. 1983, {Handbook of Stochastic Differential Equations}
  (Princeton)

\bibitem[{Giacalone {et~al.}(2000)Giacalone, Jokipii, \&
  Mazur}]{giacalone2000small}
Giacalone, J., Jokipii, J., \& Mazur, J. 2000, The Astrophysical Journal, 532,
  L75

\bibitem[{G\'{o}mez-Herrero {et~al.}(2015)G\'{o}mez-Herrero, Dresing, Klassen,
  Heber, Lario, Agueda, Malandraki, Blanco, Rodr\'{i}guez-Pacheco, \&
  Banjac}]{gomez2015}
G\'{o}mez-Herrero, R., Dresing, N., Klassen, A., {et~al.} 2015, The
  Astrophysical Journal, 799, 55, \dodoi{10.1088/0004-637x/799/1/55}

\bibitem[{{Heras} {et~al.}(1995){Heras}, {Sanahuja}, {Lario}, {Smith},
  {Detman}, \& {Dryer}}]{heras1995}
{Heras}, A.~M., {Sanahuja}, B., {Lario}, D., {et~al.} 1995, \apj, 445, 497,
  \dodoi{10.1086/175714}

\bibitem[{{Heras} {et~al.}(1992){Heras}, {Sanahuja}, {Smith}, {Detman}, \&
  {Dryer}}]{heras1992}
{Heras}, A.~M., {Sanahuja}, B., {Smith}, Z.~K., {Detman}, T., \& {Dryer}, M.
  1992, \apj, 391, 359, \dodoi{10.1086/171351}

\bibitem[{{Hu} {et~al.}(2017){Hu}, {Li}, {Ao}, {Zank}, \&
  {Verkhoglyadova}}]{Huetal2017}
{Hu}, J., {Li}, G., {Ao}, X., {Zank}, G.~P., \& {Verkhoglyadova}, O. 2017,
  Journal of Geophysical Research (Space Physics), 122, 10,938,
  \dodoi{10.1002/2017JA024077}

\bibitem[{Isenberg(1997)}]{isenberg1997hemispherical}
Isenberg, P.~A. 1997, Journal of Geophysical Research: Space Physics, 102, 4719

\bibitem[{Jokipii(1966)}]{jokipii1966cosmic}
Jokipii, J.~R. 1966, The Astrophysical Journal, 146, 480

\bibitem[{{Jokipii}(1966)}]{jokipii1966}
{Jokipii}, J.~R. 1966, \apj, 146, 480, \dodoi{10.1086/148912}

\bibitem[{{Kabin}(2001)}]{Kabin2001JPlPh..66..259K}
{Kabin}, K. 2001, Journal of Plasma Physics, 66, 259,
  \dodoi{10.1017/S0022377801001295}

\bibitem[{{Kallenrode}(1993)}]{Kallenrode1993}
{Kallenrode}, M.-B. 1993, \jgr, 98, 19037, \dodoi{10.1029/93JA02079}

\bibitem[{{Kallenrode} \& {Wibberenz}(1997)}]{KallenrodeWibberenz1997}
{Kallenrode}, M.-B., \& {Wibberenz}, G. 1997, \jgr, 102, 22311,
  \dodoi{10.1029/97JA01677}

\bibitem[{Katou \& Amano(2019)}]{katou2019}
Katou, T., \& Amano, T. 2019, The Astrophysical Journal, 874, 119,
  \dodoi{10.3847/1538-4357/ab0d8a}

\bibitem[{Kloek \& van Dijk(1978)}]{kloek1978}
Kloek, T., \& van Dijk, H.~K. 1978, Econometrica, 46, 1.
\newblock \url{http://www.jstor.org/stable/1913641}

\bibitem[{{Kozarev} {et~al.}(2013){Kozarev}, {Evans}, {Schwadron}, {Dayeh},
  {Opher}, {Korreck}, \& {van der Holst}}]{Kozarevetal2013}
{Kozarev}, K.~A., {Evans}, R.~M., {Schwadron}, N.~A., {et~al.} 2013, \apj, 778,
  43, \dodoi{10.1088/0004-637X/778/1/43}

\bibitem[{{Kwon} \& {Vourlidas}(2017)}]{kwon2017}
{Kwon}, R.-Y., \& {Vourlidas}, A. 2017, \apj, 836, 246,
  \dodoi{10.3847/1538-4357/aa5b92}

\bibitem[{Kwon {et~al.}(2014)Kwon, Zhang, \& Olmedo}]{kwon2014new}
Kwon, R.-Y., Zhang, J., \& Olmedo, O. 2014, The Astrophysical Journal, 794, 148

\bibitem[{{le Roux} \& {Webb}(2012)}]{leRouxWebb2012}
{le Roux}, J.~A., \& {Webb}, G.~M. 2012, \apj, 746, 104,
  \dodoi{10.1088/0004-637X/746/1/104}

\bibitem[{{Lee}(2005)}]{Lee2005}
{Lee}, M.~A. 2005, \apjs, 158, 38, \dodoi{10.1086/428753}

\bibitem[{{Li} {et~al.}(2003){Li}, {Zank}, \& {Rice}}]{Lietal2003}
{Li}, G., {Zank}, G.~P., \& {Rice}, W.~K.~M. 2003, Journal of Geophysical
  Research (Space Physics), 108, 1082, \dodoi{10.1029/2002JA009666}

\bibitem[{{Li} {et~al.}(2021){Li}, {Jin}, {Ding}, {Bruno}, {de Nolfo},
  {Randol}, {Mays}, {Ryan}, \& {Lario}}]{Li2021}
{Li}, G., {Jin}, M., {Ding}, Z., {et~al.} 2021, \apj, 919, 146,
  \dodoi{10.3847/1538-4357/ac0db9}

\bibitem[{{Lionello} {et~al.}(2013){Lionello}, {Downs}, {Linker},
  {T{\"o}r{\"o}k}, {Riley}, \& {Miki{\'c}}}]{lionello2013}
{Lionello}, R., {Downs}, C., {Linker}, J.~A., {et~al.} 2013, \apj, 777, 76,
  \dodoi{10.1088/0004-637X/777/1/76}

\bibitem[{{Lionello} {et~al.}(2001){Lionello}, {Linker}, \&
  {Miki{\'c}}}]{lionello2001}
{Lionello}, R., {Linker}, J.~A., \& {Miki{\'c}}, Z. 2001, \apj, 546, 542,
  \dodoi{10.1086/318254}

\bibitem[{{Luhmann} {et~al.}(2010){Luhmann}, {Ledvina}, {Odstrcil}, {Owens},
  {Zhao}, {Liu}, \& {Riley}}]{Luhmannetal2010}
{Luhmann}, J.~G., {Ledvina}, S.~A., {Odstrcil}, D., {et~al.} 2010, Advances in
  Space Research, 46, 1, \dodoi{10.1016/j.asr.2010.03.011}

\bibitem[{{Marsh} {et~al.}(2015){Marsh}, {Dalla}, {Dierckxsens}, {Laitinen}, \&
  {Crosby}}]{Marshetal2015}
{Marsh}, M.~S., {Dalla}, S., {Dierckxsens}, M., {Laitinen}, T., \& {Crosby},
  N.~B. 2015, Space Weather, 13, 386, \dodoi{10.1002/2014SW001120}

\bibitem[{{Mays} {et~al.}(2015){Mays}, {Taktakishvili}, {Pulkkinen},
  {MacNeice}, {Rast{\"a}tter}, {Odstrcil}, {Jian}, {Richardson}, {LaSota},
  {Zheng}, \& {Kuznetsova}}]{mays2015}
{Mays}, M.~L., {Taktakishvili}, A., {Pulkkinen}, A., {et~al.} 2015, \solphys,
  290, 1775, \dodoi{10.1007/s11207-015-0692-1}

\bibitem[{{Miki\`{c}} \& {Linker}(1994)}]{mikic1994}
{Miki\`{c}}, Z., \& {Linker}, J.~A. 1994, \apj, 430, 898,
  \dodoi{10.1086/174460}

\bibitem[{{Ng} \& {Reames}(1994)}]{NgReames1994}
{Ng}, C.~K., \& {Reames}, D.~V. 1994, \apj, 424, 1032, \dodoi{10.1086/173954}

\bibitem[{{Ng} {et~al.}(2003){Ng}, {Reames}, \& {Tylka}}]{Ngetal2003}
{Ng}, C.~K., {Reames}, D.~V., \& {Tylka}, A.~J. 2003, \apj, 591, 461,
  \dodoi{10.1086/375293}

\bibitem[{{Nitta} {et~al.}(2013){Nitta}, {Aschwanden}, {Boerner}, {Freeland},
  {Lemen}, \& {Wuelser}}]{nitta2013}
{Nitta}, N.~V., {Aschwanden}, M.~J., {Boerner}, P.~F., {et~al.} 2013, \solphys,
  288, 241, \dodoi{10.1007/s11207-013-0307-7}

\bibitem[{Northrop(1963)}]{northrop1963adiabatic}
Northrop, T.~G. 1963, Reviews of Geophysics, 1, 283

\bibitem[{Ontiveros \& Vourlidas(2009)}]{ontiveros2009}
Ontiveros, V., \& Vourlidas, A. 2009, The Astrophysical Journal, 693, 267,
  \dodoi{10.1088/0004-637X/693/1/267}

\bibitem[{Park {et~al.}(2013)Park, Innes, Bucik, \& Moon}]{park2013}
Park, J., Innes, D.~E., Bucik, R., \& Moon, Y.-J. 2013, The Astrophysical
  Journal, 779, 184, \dodoi{10.1088/0004-637x/779/2/184}

\bibitem[{{Parker}(1965)}]{parker1965}
{Parker}, E.~N. 1965, \planss, 13, 9, \dodoi{10.1016/0032-0633(65)90131-5}

\bibitem[{{Pesnell} {et~al.}(2012){Pesnell}, {Thompson}, \&
  {Chamberlin}}]{pesnell2012}
{Pesnell}, W.~D., {Thompson}, B.~J., \& {Chamberlin}, P.~C. 2012, \solphys,
  275, 3, \dodoi{10.1007/s11207-011-9841-3}

\bibitem[{{Prise} {et~al.}(2014){Prise}, {Harra}, {Matthews}, {Long}, \&
  {Aylward}}]{price2014}
{Prise}, A.~J., {Harra}, L.~K., {Matthews}, S.~A., {Long}, D.~M., \& {Aylward},
  A.~D. 2014, \solphys, 289, 1731, \dodoi{10.1007/s11207-013-0435-0}

\bibitem[{{Pulkkinen} {et~al.}(2010){Pulkkinen}, {Oates}, \&
  {Taktakishvili}}]{pulkkinen2010}
{Pulkkinen}, A., {Oates}, T., \& {Taktakishvili}, A. 2010, \solphys, 261, 115,
  \dodoi{10.1007/s11207-009-9473-z}

\bibitem[{{Qin} {et~al.}(2013){Qin}, {Wang}, {Zhang}, \& {Dalla}}]{qin2013}
{Qin}, G., {Wang}, Y., {Zhang}, M., \& {Dalla}, S. 2013, \apj, 766, 74,
  \dodoi{10.1088/0004-637X/766/2/74}

\bibitem[{Qin {et~al.}(2006)Qin, Zhang, \& Dwyer}]{qin2006effect}
Qin, G., Zhang, M., \& Dwyer, J. 2006, Journal of Geophysical Research: Space
  Physics, 111

\bibitem[{Qin {et~al.}(2004)Qin, Zhang, Dwyer, \&
  Rassoul}]{qin2004interplanetary}
Qin, G., Zhang, M., Dwyer, J.~R., \& Rassoul, H.~K. 2004, The Astrophysical
  Journal, 609, 1076

\bibitem[{{Reames}(2013)}]{reames2013}
{Reames}, D.~V. 2013, \ssr, 175, 53, \dodoi{10.1007/s11214-013-9958-9}

\bibitem[{{Rice} {et~al.}(2003){Rice}, {Zank}, \& {Li}}]{Riceetal2003}
{Rice}, W.~K.~M., {Zank}, G.~P., \& {Li}, G. 2003, Journal of Geophysical
  Research (Space Physics), 108, 1369, \dodoi{10.1029/2002JA009756}

\bibitem[{{Riley} {et~al.}(2001){Riley}, {Linker}, \& {Miki{\'c}}}]{riley2001}
{Riley}, P., {Linker}, J.~A., \& {Miki{\'c}}, Z. 2001, \jgr, 106, 15889,
  \dodoi{10.1029/2000JA000121}

\bibitem[{{Riley} {et~al.}(2011){Riley}, {Lionello}, {Linker}, {Mikic},
  {Luhmann}, \& {Wijaya}}]{riley2011}
{Riley}, P., {Lionello}, R., {Linker}, J.~A., {et~al.} 2011, \solphys, 274,
  361, \dodoi{10.1007/s11207-010-9698-x}

\bibitem[{{Roelof}(1969)}]{Roelof1969}
{Roelof}, E.~C. 1969, in Lectures in High-Energy Astrophysics, ed.
  H.~{{\"O}gelman} \& J.~R. {Wayland}, 111

\bibitem[{{Ruffolo}(1995)}]{Ruffolo1995}
{Ruffolo}, D. 1995, \apj, 442, 861, \dodoi{10.1086/175489}

\bibitem[{{Scherrer} {et~al.}(2012){Scherrer}, {Schou}, {Bush}, {Kosovichev},
  {Bogart}, {Hoeksema}, {Liu}, {Duvall}, {Zhao}, {Title}, {Schrijver},
  {Tarbell}, \& {Tomczyk}}]{scherrer2012}
{Scherrer}, P.~H., {Schou}, J., {Bush}, R.~I., {et~al.} 2012, \solphys, 275,
  207, \dodoi{10.1007/s11207-011-9834-2}

\bibitem[{{Schlickeiser}(2002)}]{schlickeiser2002}
{Schlickeiser}, R. 2002, {Cosmic Ray Astrophysics}

\bibitem[{{Skilling}(1971)}]{skilling1971}
{Skilling}, J. 1971, \apj, 170, 265, \dodoi{10.1086/151210}

\bibitem[{{Thompson}(1962)}]{Thompson1962}
{Thompson}, W.~B. 1962, {An Introduction to Plasma Physics} (Addison Wesley),
  86--95

\bibitem[{{Verdini} \& {Velli}(2007)}]{verdini2007}
{Verdini}, A., \& {Velli}, M. 2007, \apj, 662, 669, \dodoi{10.1086/510710}

\bibitem[{Zank {et~al.}(2000)Zank, Rice, \& Wu}]{zank2000particle}
Zank, G., Rice, W., \& Wu, C. 2000, Journal of Geophysical Research: Space
  Physics, 105, 25079

\bibitem[{{Zank} {et~al.}(1996){Zank}, {Matthaeus}, \& {Smith}}]{zank1996}
{Zank}, G.~P., {Matthaeus}, W.~H., \& {Smith}, C.~W. 1996, \jgr, 101, 17093,
  \dodoi{10.1029/96JA01275}

\bibitem[{{Zhang}(2000)}]{zhang2000}
{Zhang}, M. 2000, \apj, 541, 428, \dodoi{10.1086/309429}

\bibitem[{{Zhang}(2006)}]{zhang2006}
---. 2006, Journal of Geophysical Research (Space Physics), 111, A04208,
  \dodoi{10.1029/2005JA011323}

\bibitem[{Zhang {et~al.}(2009)Zhang, Qin, \& Rassoul}]{zhang2009propagation}
Zhang, M., Qin, G., \& Rassoul, H. 2009, The Astrophysical Journal, 692, 109

\bibitem[{{Zhang} \& {Zhao}(2017)}]{ZhangZhao2017}
{Zhang}, M., \& {Zhao}, L. 2017, \apj, 846, 107,
  \dodoi{10.3847/1538-4357/aa86a8}

\bibitem[{Zhao \& Zhang(2018)}]{zhao2018}
Zhao, L., \& Zhang, M. 2018, The Astrophysical Journal Letters, 859, l29,
  \dodoi{10.3847/2041-8213/aac6cf}

\bibitem[{{Zuo} {et~al.}(2013{\natexlab{a}}){Zuo}, {Zhang}, \&
  {Rassoul}}]{zuoetal2013a}
{Zuo}, P., {Zhang}, M., \& {Rassoul}, H.~K. 2013{\natexlab{a}}, \apj, 776, 93,
  \dodoi{10.1088/0004-637X/776/2/93}

\bibitem[{{Zuo} {et~al.}(2013{\natexlab{b}}){Zuo}, {Zhang}, \&
  {Rassoul}}]{zuoetal2013b}
---. 2013{\natexlab{b}}, \apj, 767, 6, \dodoi{10.1088/0004-637X/767/1/6}

\end{thebibliography}
\bibliographystyle{aasjournal}
\end{document}